\documentclass{emulateapj}
\usepackage{amsfonts,amssymb,rotating,lscape,graphics}

\newcommand{\si}[1]{\ensuremath{_{\scriptsize{\textrm{#1}}}}}

\begin{document}
\title{A \emph{Chandra} Look at Five of the Broadest Double-Peaked Balmer-Line Emitters}

\author{
Iskra V. Strateva\altaffilmark{1},
W. N. Brandt\altaffilmark{2},
Michael Eracleous\altaffilmark{2,3},
Gordon Garmire\altaffilmark{2}
}

\altaffiltext{1}{Max Planck Institute for Extraterrestrial Physics, Garching, 85741, Germany}
\altaffiltext{2}{Department of Astronomy and Astrophysics, 525 Davey Lab, 
The Pennsylvania State University, University Park, PA 16802} 
\altaffiltext{3}{Center for Gravitational Wave Physics, The Pennsylvania State University, University Park, PA 16803}

\addtocounter{footnote}{2} 

\begin{abstract}
We study the 0.5--10\,keV emission of a sample of five of the broadest double-peaked Balmer-line emitters with \emph{Chandra}. The Balmer lines of these objects originate close (within a few hundred gravitational radii) to the central black holes of the Active Galactic Nuclei (AGNs), and their double-peaked profiles suggest an origin in the AGN accretion disk. We find that four of the five targets can be modeled by simple power-law continua with photon indices ($1.6\leq\Gamma\leq1.8$) typical of similar luminosity AGNs. One object, SDSS\,J0132$-$0952, shows evidence of ionized intrinsic absorption. The most-luminous SDSS double-peaked emitter, SDSS\,J2125$-$0813, has either an unusual flat spectrum ($\Gamma\sim1$) or is also highly absorbed. It is the only double-peaked emitter for which no external illumination is necessary to account for the Balmer line emission. The strength of the Balmer-line emission in the remaining four objects suggests that the total line flux likely exceeds the viscous energy that can be extracted locally from the accretion disk and external illumination is necessary. All five double-peaked emitters have unusually strong X-ray emission relative to their UV/optical emission, which is the likely source of the external illumination necessary for the production of the observed strong broad lines. On average about 30\% of their bolometric luminosities are emitted between 0.5--10\,keV. The spectral energy distributions of the five double-peaked emitters show the big blue bumps characteristic of radiatively efficient accretion flows. The Balmer line profiles, as well as the optical and X-ray fluxes of the double-peaked emitters, are highly variable on timescales of months to years in the AGN rest frame.
\end{abstract}    


\section{INTRODUCTION}
\label{intro}

Double-peaked emitters are Active Galactic Nuclei (AGNs) whose  broad Balmer-line profiles have a characteristic ``double-humped'' shape, with one peak blueward of the rest-frame wavelength of the line and another one redward of it. The best explanation for the origin of the double-peaked shape involves emission from a region in the AGN accretion disk. Since the first examples of double-peaked emission lines in AGNs were published in the 1980s \citep[e.g.,][]{Oke87,CH89a}, more than 100 double-peaked emitters have been found. Some of these objects have also been studied in the X-ray band \citep[e.g.,][]{2Pxray1,2Pxray2,2Pxray3,2Pxray4,S06}, but most of these studies focused on individual objects or on serendipitous X-ray detections of limited quality.

\begin{figure}
\plotone{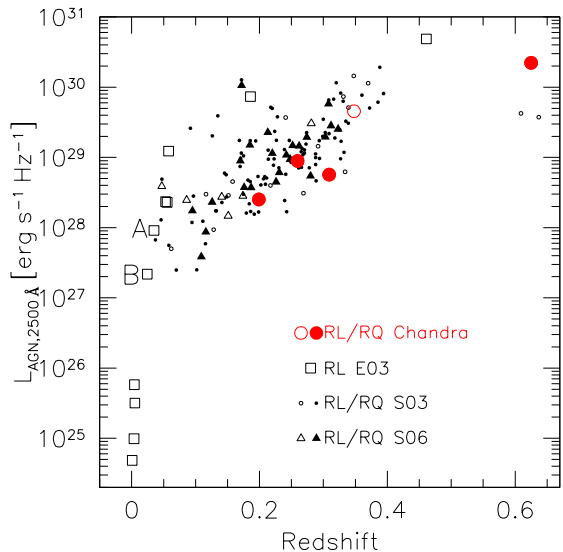}
\caption{The redshift-luminosity distribution of the five-object \emph{Chandra} sample with respect to the full S03 sample, the \emph{ROSAT} sample studied by S06, and the X-ray observed E03 sample. Two of the E03 objects, discussed in \S~\ref{intro}, are denoted by letters:  A stands for Pictor\,A, and B stands for Arp\,102B. All monochromatic luminosities are corrected for host-galaxy contamination.
\label{lzplot}}
\end{figure}

\begin{deluxetable}{ccccc} 
\tablewidth{0pt} 
\tablecaption{\emph{Chandra} Sample}
\tablehead{\colhead{SDSS Name} &\colhead{$\log R$} &\colhead{$z$} &\colhead{FWHM} &\colhead{$M_i$} \\
\colhead{(1)} &\colhead{(2)} &\colhead{(3)} &\colhead{(4)} &\colhead{(5)} \\
\colhead{HHMMSS.ss$\pm$DDMMSS.s} &\colhead{} &\colhead{} &\colhead{km\,s$^{-1}$}&\colhead{mag}}
\startdata                                                                                                      
J013253.31$-$095239.5  & $<0.7$ & 0.260 & 15200 & $-22.9$\\   
J100443.44$+$480156.5 & $<0.6$ & 0.199 & 16000 & $-22.8$\\   
J123807.77$+$532556.0 & $2.3$ & 0.348  & 15400 & $-24.4$\\   
J140720.70$+$023553.1 & $<1$  & 0.309  & 20100 & $-22.9$\\   
J212501.21$-$081328.6  & $<0.2$ & 0.625 & 14000 & $-26.8$ \\ 
\enddata
\tablecomments{(1) SDSS name given in the J2000 epoch RA and Dec form; (2) Radio loudness parameter; (3) Redshift; (4) Full-Width at Half Maximum of the H$\alpha$ (objects \# 1 through 4) or H$\beta$ (object \# 5) line; (5) Extinction- and K-corrected \citep[assuming a power-law continuum with index $\alpha_{\nu}=-2.45$ for \hbox{$\lambda>5000$\,\AA};][]{Berk} SDSS $i$-band model magnitude (including the host-galaxy contribution).}
\label{tabS}
\end{deluxetable}

\begin{deluxetable*}{ccccccccc}  
\tablewidth{0pt} 
\tabletypesize{\small}
\tablecaption{X-ray Observations} 
\tablehead{\colhead{Object} &\colhead{ObsID} &\colhead{Instrument} &\colhead{ObsDate} &\colhead{$t_{\textrm{exp}}$} &\colhead{Counts} & \colhead{Count Rate} &\colhead{PSF FWHM}&\colhead{\emph{ROSAT}}\\ 
\colhead{(1)} &\colhead{(2)} &\colhead{(3)} &\colhead{(4)} &\colhead{(5)} &\colhead{(6)} &\colhead{(7)} &\colhead{(8)}&\colhead{(9)}\\
\colhead{HHMM$\pm$DDMM} &\colhead{} &\colhead{} &\colhead{yyyy-mm-dd} &\colhead{sec} &\colhead{}&\colhead{counts\,s$^{-1}$} &\colhead{$''$} &\colhead{counts\,s$^{-1}$}}
\startdata
J0132$-$0952& 6770 & ACIS-7             & 2006-06-24 & 8151 & 727 & $0.089\pm0.003$ & $0.85\pm0.03$ & $\lesssim0.011$\\
J1004$+$4801 & 6771 & ACIS-235678 & 2006-01-31 & 8628 & 359 & $0.042\pm0.002$ & $0.80\pm0.03$ & $0.04\pm0.01$\\
J1238$+$5325 & 6773 & ACIS-7           & 2006-05-31 & 4061 & 1133 & $0.279\pm0.008$ & $0.75\pm0.02$ & $0.10\pm0.02$\\
J1407$+$0235 & 6772 & ACIS-235678 & 2005-12-07 & 5471 & 1348* & $0.246\pm0.007$* & $1.1\pm0.1$ & $\lesssim0.014$\\
J2125$-$0813 & 6774 & ACIS-7             & 2006-04-02 & 4110 & 155   & $0.038\pm0.003$ & $1.1\pm0.1$ & $0.06\pm0.01$\\
\enddata
\tablecomments{The asterisk (*) denotes observed values, which are not corrected for pile-up. (1) Short object name; (2) \emph{Chandra} Observation ID; (3) ACIS chips used; all observations used the 1/8 subarray except ObsID 6772, which used the 1/2 subarray; (4) Date of the observation; (5) Effective exposure time, $t_{\textrm{exp}}$; (6) Total counts in the \hbox{0.3--10\,keV} spectrum; (7) Count rate for the ACIS-S3 chip; (8) The Full Width at Half Maximum (FWHM) of the ACIS image Point Spread Function (PSF); (9) The \hbox{0.5--2\,keV} \emph{ROSAT} count rate.}
\label{tab1}
\end{deluxetable*}

The small sample of double-peaked emitters studied in detail in the X-ray band to date consists of predominantly low-luminosity radio-loud\footnote{Following \citet{rl} we define an AGN as radio loud if $\log_{10} R=\log[F\si{20\,cm}/F_i]=0.4(i-t)>1$, where $i$ is the Sloan Digital Sky Survey $i$-band magnitude and $t=-2.5\log(F\si{20\,cm}/3136\,\textrm{Jy})$ is the 20\,cm AB radio magnitude.} (RL) objects -- broad line radio galaxies (BLRGs, Eracleous et al.~1994, 2003, hereafter E03). Based on the prototype disk emitter (Arp~102B) and the small original BLRG sample, a model was developed to explain the existence of double-peaked emission in AGNs. \citet{CH89b} proposed that BLRGs and other low-accretion-rate, low-luminosity\footnote{$L\si{bol}<1.3\times10^{43}\textrm{\,erg\,s}^{-1}=10^{-3} \eta M_8^{-1}L\si{Edd}$, where $M_8$ is the black hole mass in units of $10^8$\,$M_{\sun}$, $\eta$ is the accretion efficiency, and $L\si{Edd}$ is the Eddington luminosity. Efficient accretors can convert up to $\eta\lesssim0.3$ \citep[for a maximally rotating black hole;][]{Thorne74} of the gravitational energy into radiation.} AGNs -- which do not accrete through a standard thin disk -- could exhibit low-ionization lines with double-peaked profiles. The line profiles are produced as the outer portions of the disk are illuminated by an X-ray-emitting elevated structure in the center -- a structure that is characteristic of radiatively inefficient accretion flows \citep[RIAFs; e.g,][]{Rees82,adaf,adios,cdaf} and replaces the standard thin disk. The illumination provided by the central elevated structure in AGNs with RIAFs is necessary, but not sufficient, for the production of double-peaked lines of the observed strength: all BLRGs with double-peaked lines require external illumination (since the viscous power available locally in the disk is insufficient to explain the line strength), but not all BLRGs or other RIAF AGNs have double-peaked lines. The association of RL AGNs with RIAFs could explain the higher frequency of double-peaked emitters observed among BLRGs (20\%; E03), compared to 3\% of low-redshift Sloan Digital Sky Survey (SDSS) AGNs. Recent work by \citet{Lewis}, however, has shown that not all BLRG double-peaked emitters with low optical luminosities are inefficient emitters: Pictor~A, for example, emits a bolometric luminosity of more than 8\% of  the Eddington luminosity, which is too high to support a RIAF \citep[RIAF solutions exist for $\dot M\si{crit}=\alpha^2\dot M\si{Edd}<0.09\dot M\si{Edd}$, for a viscosity parameter $\alpha<0.3$, and the radiated energy is further reduced by the accretion efficiency;][]{NY95}.

\begin{figure*}
\plotone{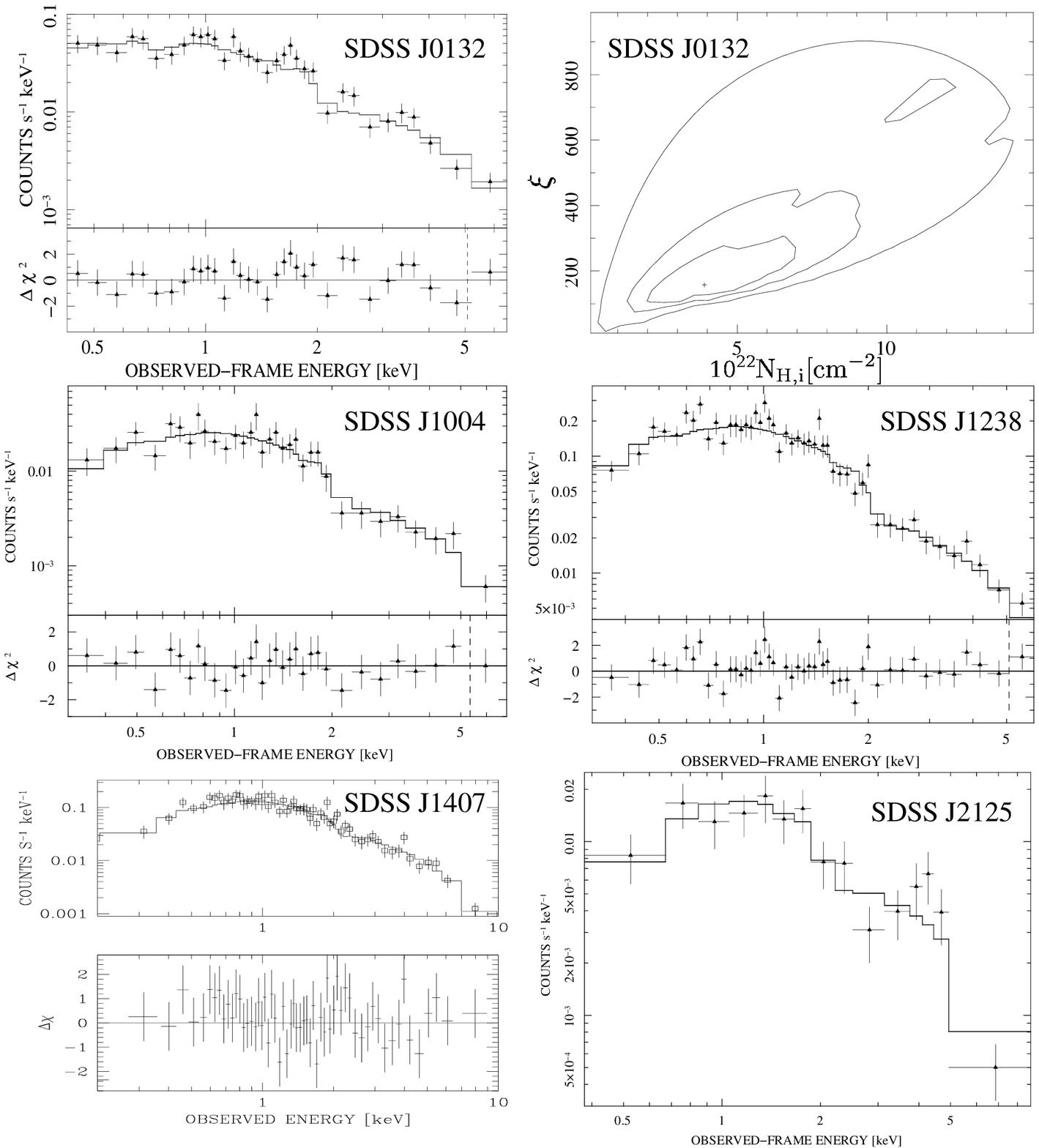}
\caption{X-ray spectra of the five double-peaked emitters overlaid with the best-fit models. The panels are arranged in order of increasing RA, with the RA increasing to the right and down. The second panel displays the $1-3\sigma$ confidence contours for the gas column density and ionization parameter for the SDSS\,J0132$-$0952 fit shown in the first panel. The SDSS\,J1407$-$0235 fit takes into account the pile-up estimated to be 12\%. The last panel gives the SDSS\,J2125$-$0813 Cash-statistic fit which has no $\Delta\chi^2$ residuals; $\Delta\chi$ refers to the residuals in units of $\sigma$ with errorbars of size 1.
\label{Xspec}} 
\end{figure*}

The systematic search for AGNs with double-peaked Balmer lines in the SDSS revealed that as many as 3\% of $z<0.33$ objects are double-peaked emitters, regardless of their radio loudness (Strateva et al.~2003; hereafter, S03). Moreover, a handful of luminous higher-redshift double-peaked emitters were discovered serendipitously based on their H$\beta$ and \ion{Mg}{2} line profiles. The expansion of the parameter space of the double-peaked emission-line AGNs to include radio-quiet (RQ, $\log R\leq 1$) and higher luminosity objects, as well as the discovery of low optical luminosity objects like Pictor~A that are efficient accretors,  require a critical evaluation of the existing models of illumination and the conditions necessary for the existence of double-peaked emission lines.

X-ray observations of the new extended sample are particularly important, since they allow us to obtain more accurate estimates of the bolometric luminosities of the sources and to gauge the need for extra illumination of the accretion disk. Double-peaked emitters which are efficient accretors likely require an alternative mode of illumination. The development of new scenarios can be guided by observations in the X-ray band. Basic X-ray absorption constraints, for example, provide sensitive probes of the nuclear conditions: (1) they allow us to see any intervening material right up to the AGN core, and (2) they are sensitive to molecular, atomic, or ionized gas and dust. The \hbox{X-ray} absorption constraints  could provide sensitive tests of specific models proposed to explain the expanded population of double-peaked emitters. A recent model by  \citet{jet}, for example, aims to explain the disk illumination necessary for the existence of the double-peaked lines with the presence of jets in RL, and slower, less-collimated outflows in RQ double-peaked emitters. The predictions of this model can be tested directly by obtaining X-ray spectroscopy of double-peaked emitters and looking for signs of these outflows in absorption. 

In addition to testing specific models of illumination, comparison of the X-ray emission of double-peaked emitters to that of typical AGNs is essential for uncovering the necessary and sufficient conditions for the dominance of accretion-disk Balmer-line emission in this type of AGN. Such findings, which will also elucidate the physical conditions in AGNs with the broadest  observed Balmer lines, will have implications for the (still uncertain) structure of the broad line region (BLR) in AGNs in general. It is conceivable that the accretion disk is part of the BLR in all AGNs, but its contribution dominates the Balmer-line emission only in the small subsample of AGNs with the broadest lines \citep[see also][for some tantalizing ideas about the structure of the BLR]{Laor07}.

In this paper we present the first results of an exploratory campaign to map the X-ray properties of a representative sample of AGNs with the broadest double-peaked Balmer-lines. Using the \emph{Chandra} X-ray Observatory we obtained high-quality X-ray images and spectra for five sources, complemented by contemporaneous optical spectra from the ground. Our five targets span the redshift and optical luminosity plane of the SDSS double-peaked sample presented in S03. The paper is organized as follows: in \S\ref{sample} we present the current sample and its selection criteria, followed by the analysis of the X-ray and optical observations of each individual source in \S\S\ref{analysis} and \ref{optspec}, and the discussion and conclusions in \S\ref{conclusions}.  Throughout this paper we use the \citet{Spergel07} concordance cosmology and quote all flux and luminosity errors as 90\% confidence limits; all other errors quoted are 1$\sigma$ Gaussian equivalent.

\begin{figure*}
\plotone{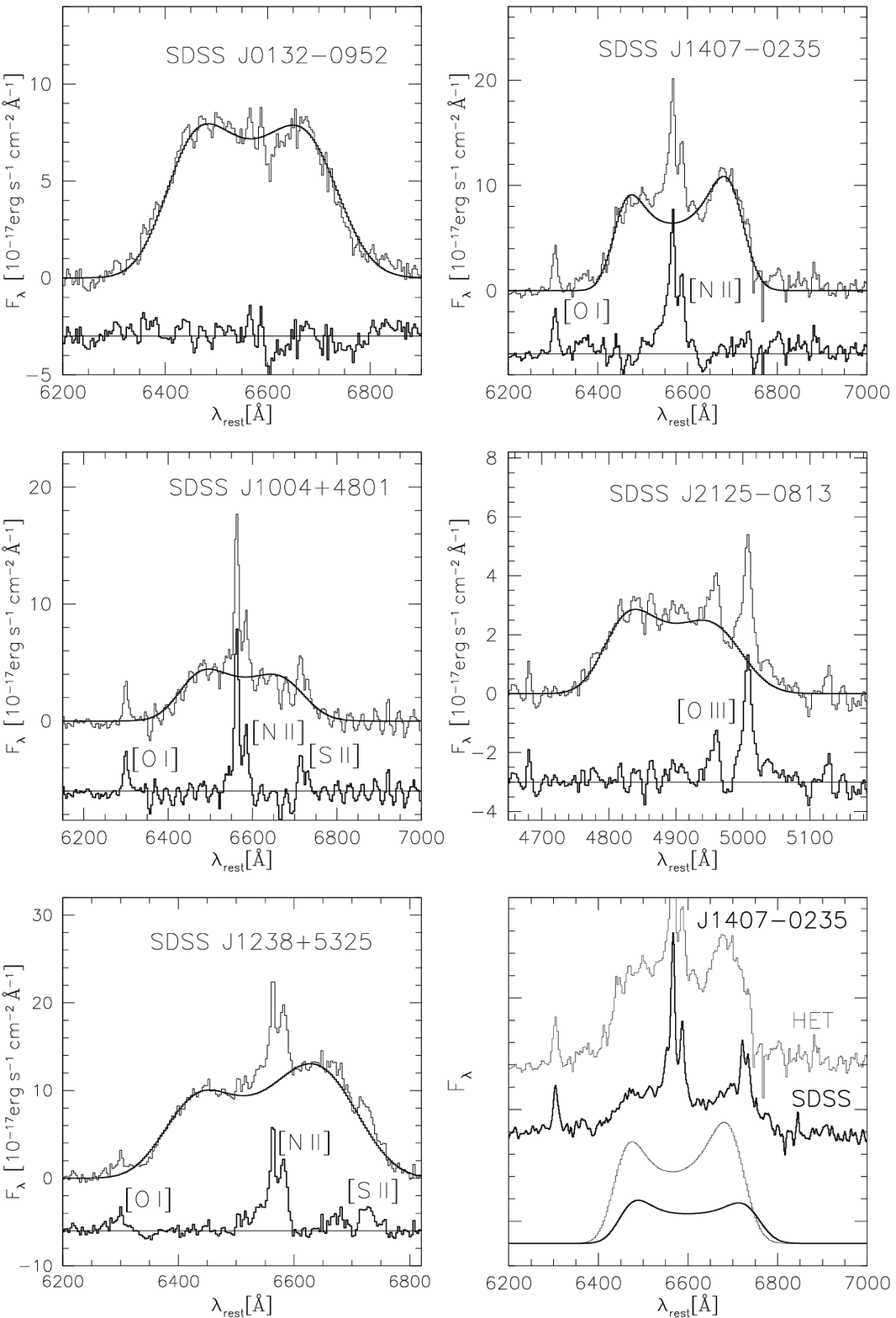}
\caption{The H$\alpha$ (first four) and H$\beta$ (middle right panel) line profiles of the five double-peaked emitters observed with \emph{Chandra} overlaid with the best-fit accretion-disk emission models. The panels are arranged in order of increasing RA, with the RA increasing down and to the right. Each panel shows the fit residuals, including the indicated narrow lines, displaced to negative values for clarity. The bottom right panel shows the H$\alpha$ variability of one of our targets, SDSS\,J1407$-$0235. The top curves denote the HET and SDSS spectra, while the bottom two curves show the accretion-disk fits; for clarity, the continuum levels have been displaced by arbitrary values in all cases.
\label{Ospec}}
\end{figure*}

\section{SAMPLE SELECTION}
\label{sample}

The sample of five SDSS double-peaked emitters observed with \emph{Chandra} was selected to span the part of the luminosity-redshift-RL space which was not covered by the initial BLRG sample. Figure~\ref{lzplot} shows the positions of the five targets on the redshift vs.\ 2500\,\AA\ monochromatic-luminosity plane, using the rest-frame flux densities measured during the SDSS epoch. The full SDSS sample of double-peaked emitters from S03 is also shown for comparison, as well as the sample of SDSS double-peaked emitters serendipitously detected with \emph{ROSAT} and presented in Strateva et al.~(2006; hereafter, S06). The \emph{Chandra} sample is representative of the full SDSS double-peaked emitter sample with regard to optical luminosity, redshift, and radio-loudness (to the extent that most SDSS double-peaked emitters are RQ). The prototype disk emitter, Arp\,102\,B, is fainter at 2500\,\AA\ than most SDSS double-peaked emitters, as is the optically-brighter, efficient accretor, Pictor\,A.

The initial BLRG sample of \citet{E94,E03} has been extended with serendipitously discovered double-peaked emitters in a handful of low-ionization nuclear emission-line region galaxies (LINERs) and quasars. A small number of sources from this original sample of double-peaked emitters have serendipitous or pointed X-ray coverage; in all but 9 cases only soft-band coverage by \emph{ROSAT} is available, with too few counts for spectral fits. Among the 9 double-peaked emitters from the original sample with \hbox{0.3--10\,keV} \emph{XMM-Newton} and \emph{Chandra} observations, 4 are LINERs (NGC\,1097, M\,81, NGC\,4203, NGC\,4579; S06 and references therein)  and the remaining 5 are BLRGs (Arp\,102\,B, 3C\,390.3, Pictor\,A, PKS\,0921$-$213, and 3C\,382; S06, Grandi et al.~2006). The X-ray data for these 9 double-peaked emitters are summarized in \citet{S06}.  These are predominantly low UV luminosity and redshift ($z<0.1$) objects, as shown in Figure~\ref{lzplot}. The new \emph{Chandra} targets were selected to be predominantly RQ (four out of the five), higher luminosity and redshift (\hbox{$0.199 \leq z \leq 0.625$}) counterparts to the original-sample sources studied in the hard X-ray band, forming an ideal comparison sample of double-peaked emitters. The selection criteria used only optical data, and the sample is therefore unbiased with respect to the X-ray properties of the selected targets. 
We present the five targets in Table~\ref{tabS}, listing the SDSS $i$-band magnitudes, redshifts and radio-loudness, as well as the Full-Widths at Half Maximum (FWHMs) of the H$\alpha$ (or H$\beta$) lines. 

The five double-peaked emitters observed with \emph{Chandra} are among the 16 AGNs with the broadest published Balmer lines: \hbox{$\textrm{FWHM}>14000$\,km\,s$^{-1}$}. Among the parent sample of over 3000 SDSS AGNs from which the SDSS double-peaked emitters were selected, virtually all AGNs with such extremely broad Balmer lines showed line profiles indicative of a disk origin. Thus, with regard to their Balmer line widths, the five \emph{Chandra} targets are not representative of the full sample of SDSS double-peaked emitters. They were selected because the widths of their Balmer lines suggest an origin very close to the center of the disk. As such they are the most suitable objects for addressing the question of the energy budget of the line-emitting portion of the disk. By extension, this set of objects can also help us address the important question of why only a small fraction of AGNs show double-peaked Balmer line profiles, while the majority of the profiles are single peaked.

\section{\emph{Chandra} Imaging Spectroscopy}
\label{analysis}

Table~\ref{tab1} summarizes the observations carried out during \emph{Chandra} Cycle~8. The exposure times were selected to yield enough counts for a basic spectroscopic fit in each case ($>$100 counts). Three of the targets (SDSS\,J1004$+$4801, SDSS\,J1238$+$5325, and SDSS\,J2125$-$0813) had prior serendipitous \emph{ROSAT} detections, while the remaining two have \emph{ROSAT} all sky survey (RASS) upper limits. We used CIAO v.3.4 and XSPEC v.12.3.0 to analyze the data, starting with the event\,1 files. We fit all spectra with more than 200 counts in the  0.3--10\,keV band  using $\chi^2$-minimization and use the Cash statistic ($C$-stat) for the one object with less than 200 counts.  The results for each of the five double-peaked emitters are summarized below, where we also discuss their X-ray variability.

\subsection{SDSS\,J0132$-$0952}
\label{spec6770}

SDSS\,J0132$-$0952 is a RQ double-peaked emitter which was observed with \emph{Chandra} for 8.2\,ks.  The X-ray image of the target is a point source with a Point Spread Function (PSF) FWHM~$=0.85\pm0.03''$ (Table~\ref{tab1}). Based on the SDSS image, our target lies in a cluster at \hbox{$z=0.26$} identified by \citet{cluster}; it is situated at a projected distance of $\sim230$\,kpc from the brightest cluster galaxy.

The \hbox{0.3--10\,keV} spectrum of SDSS\,J0132$-$0952 cannot be modeled  by a simple power law absorbed by neutral hydrogen with a column density equal to the Galactic value ($\chi^2/DoF=50/32$, where $DoF$ stands for the number of degrees of freedom). Since intrinsic absorption is common among AGNs, we attempted to model the spectrum by adding an intrinsic absorber. The addition of a neutral intrinsic absorber does not improve the fit significantly ($\chi^2/DoF=49/31$), because the spectrum is too soft; allowing for partial coverage improves the fit, but the model is only marginally acceptable ($\chi^2/DoF=42/30$). The fit residuals are suggestive of an absorption edge around \hbox{0.7--0.8\,keV}, and we attempted to fit the spectrum with a ionized absorber model  \citep[\emph{absori} in XSPEC;][]{Done92}. The spectral fit is shown in Fig.~\ref{Xspec}. The fit is improved at $>98$\% confidence ($\chi^2/DoF=37/30$); the unabsorbed spectrum is a power law with $\Gamma=1.70\pm0.08$, typical for similar X-ray luminosity AGNs, which have an average photon index of $\left<\Gamma\right>=1.74\pm0.09$ \citep[e.g.,][]{piconcelli}. The column density of the ionized absorber is constrained to $N\si{H,i}=(4\pm2)\times10^{22}$\,cm$^{-2}$ with an ionization parameter\footnote{The ionization state is defined by the ionization parameter, $\xi$, with $\xi=L/(nR^2)$, where $L$ denotes the luminosity of the source, $R$ is the distance to the source, and $n$ is the Hydrogen column density \citep{T69}.} $\xi=160\pm70$\,erg\,s$^{-1}$\,cm. The \emph{absori} fit parameters are given in Table~\ref{tab2} and the estimated fluxes and monochromatic luminosities in Table~\ref{tab3}. We show the $1-3\sigma$ contours for two interesting parameters in Figure~\ref{Xspec}. Since the XSPEC \emph{absori} model is not fully self consistent (it treats the temperature and ionization parameter as independent parameters, for example), we used the \emph{etable} XSTAR model \citep{K82,K04}, as implemented in XSPEC,\footnote{See http://heasarc.gsfc.nasa.gov/docs/software/xstar/xstar.html for more details.} to verify the \emph{absori} model parameters. The ionization parameter\footnote{Note that the luminosity ranges used in the definition of the ionization parameter are chosen differently in different photoionization codes. The XSPEC \emph{absori} model computes the ionizing luminosity between 5\,keV and 300\,keV, compared to 1\,Ry to 1000\,Ry (which corresponds to 13.6\,eV to 13.6\,keV) in the XSTAR models. This results in a difference in the ionization parameter $\xi$ of about a factor of two, with  $\xi_{absori}\sim 2 \xi\si{XSTAR}$.} and absorber column density are both consistent: $N\si{H,i}=3_{-3}^{+10}\times10^{22}$\,cm$^{-2}$ and $\xi\si{XSTAR}=250_{-150}^{+550}$\,erg\,s$^{-1}$\,cm, but the XSTAR model constraints are much looser than those obtained with the XSPEC \emph{absori} model. This is a result of the limited signal-to-noise ratio of the current data, for which the \emph{absori} model, which has no sophisticated line information, is sufficiently detailed. 

SDSS\,J0132$-$0952 falls within RASS observation RS931905N00 (June 1990), but is not detected in 440\,s. We estimate a count-rate upper limit of $\lesssim$0.011\,s$^{-1}$ in the \hbox{0.5--2\,keV} band (see Table~\ref{tab1}), corresponding to an expected \emph{Chandra} ACIS-S count rate of $\lesssim$\,0.036\,s$^{-1}$, fainter than the observed 0.5--2\,keV ACIS-S count rate of $0.056\pm0.003$\,s$^{-1}$. This suggests that the source is variable, brightening by at least $\sim$60\% in the last 17 years (corresponding to $\sim13.5$\,years in the source rest-frame).

\subsection{SDSS\,J1004$+$4801}
\label{spec6771}

\begin{deluxetable*}{ccccccc}
\tablewidth{0pt} 
\tablecaption{X-ray Spectral Fits} 
\tablehead{\colhead{Object} &\colhead{$\Gamma$} &\colhead{$N\si{H,G}$} &\colhead{$N\si{H,i}$}  &\colhead{$EW\si{Fe\,K$\alpha$}$} &\colhead{$\chi^2/DoF$} & \colhead{Model} \\ 
\colhead{(1)} &\colhead{(2)} &\colhead{(3)} &\colhead{(4)} &\colhead{(5)} &\colhead{(6)} &\colhead{(7)}\\
\colhead{HHMM$\pm$DDMM} &\colhead{} &\colhead{cm$^{-2}$} &\colhead{cm$^{-2}$} &\colhead{keV} &\colhead{} &\colhead{}}
\startdata
J0132$-$0952 & $1.70\pm0.08$ & $3.0\times10^{20}$ & $(4\pm2)\times10^{22}$ & $<0.5$ & 37/30 & \emph{wabs(absori*zpow)} \\
& & &  $\xi=160\pm70$ & & &\\
J1004$+$4801 & $1.60\pm0.09$ & $9.0\times10^{19}$ & ...  & $<0.9$ & 21/32 & \emph{wabs*zpow} \\
J1238$+$5325 & $1.77\pm0.05$ & $1.6\times10^{20}$ & ... & $<0.3$ & 54/46 & \emph{wabs*zpow} \\
J1407$+$0235 & $1.60\pm0.05$ & $2.4\times10^{20}$ & ... & ... &  49/58 & \emph{jdpileup(abs*power)} \\
J2125$-$0813 &  $1.0\pm0.1$& $5.3\times10^{20}$ & ... & $\sim0.7?$ & 123/118 & \emph{wabs*zpow}\\
\enddata
\tablecomments{(1) Short object name; (2) Power-law photon index; (3) Galactic hydrogen column density; (4) Intrinsic absorbing column density $N\si{H,i}$ and ionization parameter of the absorbing gas, $\xi$ in units of \,erg\,s$^{-1}$\,cm, whenever appropriate; (5) Fe\,K$\alpha$ line equivalent width or upper limit; (6) $\chi^2$ (first 4 rows) or $C$-stat (last row) of the model fit and the number of degrees of freedom, $DoF$; (7) XSPEC/SHERPA model.}
\label{tab2}
\end{deluxetable*}

SDSS\,J1004$+$4801 is a RQ double-peaked emitter that was observed with \emph{Chandra} for 8.6\,ks (Table~\ref{tab1}).  The \hbox{0.3--10\,keV} spectrum has 359 counts and can be modeled with a  simple power law modified by absorption equal to the Galactic value ($\chi^2/DoF=21/31$; see Table~\ref{tab2} and Figure~\ref{Xspec}). We measure a power-law photon index of $\Gamma=1.60\pm0.09$, which is rather flat, but within the range observed for RQ AGNs; the soft-band flux is $F\si{\hbox{0.5--2\,keV}}=(1.4\pm0.2)\times10^{-13}$\,ergs\,cm$^{-2}$\,s$^{-1}$. Considering the small number of counts in the spectrum, we can alternatively fix the power-law slope of the spectral fit to the average value observed in similar redshift and luminosity AGNs, $\Gamma=1.74$ \citep[e.g.,][]{piconcelli}, and allow for some intrinsic absorption. The model fit including a neutral absorber is acceptable ($\chi^2/DoF=22/31$), but the inferred intrinsic absorbing column density is small, $N\si{H,i}\lesssim6\times10^{20}$\,cm$^{-2}$. 

SDSS\,J1004$+$4801 was detected in the RASS in October 1990, with a 0.5--2\,keV count rate of $0.04\pm0.01$\,s$^{-1}$ and a hardness ratio\footnote{The hardness ratio used here is defined as $(A - B)/(A + B)$, where $A$ refers to the \emph{ROSAT} PSPC count rate in the hard (0.5--2.0\,keV) and B in the soft (0.1--0.4\,keV) band.} of 0.01 (corresponding to $\Gamma\approx1.7$ with absorption equal to the Galactic value). Assuming the \emph{Chandra} power-law model slope and absorption equal to the Galactic value, the \emph{ROSAT} soft-band flux was $F\si{\hbox{0.5--2\,keV}}=(4.4\pm1.1)\times10^{-13}$\,ergs\,cm$^{-2}$\,s$^{-1}$. This is significantly brighter than the 2006 \emph{Chandra} soft-band flux given above.\footnote{Restricting the \emph{Chandra} spectral fit to the \emph{ROSAT} band does not affect the result.}

\subsection{SDSS\,J1238$+$5325}
\label{spec6773}

SDSS\,J1238$+$5325 (Table~\ref{tab1}) is a RL double-peaked emitter which we observed with \emph{Chandra} using the ACIS-S 1/8 subarray. The 0.3--10\,keV spectrum can be represented by a simple power law with photon index $\Gamma=1.77\pm0.05$ modified by Galactic absorption (see Table~\ref{tab2} and Figure~\ref{Xspec}). Any intrinsic neutral absorption is constrained to be $N\si{H,i}<3\times10^{21}$\,cm$^{-2}$.

SDSS\,J1238$+$5325 was previously detected with \emph{ROSAT} in December 1990. Assuming that the \emph{Chandra} power-law model modified by Galactic absorption applies, the soft-band flux was  $F\si{\hbox{0.5--2\,keV}}=(1.2\pm0.2)\times10^{-12}$\,ergs\,cm$^{-2}$\,s$^{-1}$ during the \emph{ROSAT} observation, essentially unchanged from the 2006 \emph{Chandra} observation. 

\subsection{SDSS\,J1407$+$0235}
\label{spec6772}

SDSS\,J1407$-$0235 is a RQ double-peaked emitter observed by \emph{Chandra} for 5.5\,ks (Table~\ref{tab1}). The source falls within two RASS $\sim$390\,s pointings: RS931738N00 and RS931638N00 observed on 1991 August 13. The estimated count-rate upper limits, $\lesssim0.014$\,s$^{-1}$, are consistent  for the two observations, corresponding to an expected \emph{Chandra} ACIS-S count rate of $\lesssim0.066$\,s$^{-1}$ (\hbox{0.3--10\,keV}). The actual observed count rate (using the 1/2 subarray) was $0.25\pm0.01$\,s$^{-1}$ (a brightening by at least a factor of 4) and the source is piled up. We used SHERPA's  \emph{jdpileup} model to estimate the pile-up and model the X-ray spectrum. The estimated pile-up fraction is 12\%, and the spectrum can be modeled by a simple power law with photon index $\Gamma=1.60\pm0.05$ modified by Galactic absorption.  The pile-up and absorption corrected fluxes in the hard and soft bands are $F\si{2-10\,keV}=1.7\times10^{-12}$\,erg\,cm$^{-2}$\,s$^{-1}$ and $F\si{0.5-2\,keV}=8.1\times10^{-13}$\,erg\,cm$^{-2}$\,s$^{-1}$, respectively (see Table~\ref{tab3}).

\subsection{SDSS\,J2125$-$0813}
\label{spec6774}

SDSS\,J2125$-$0813  is the highest optical luminosity SDSS double-peaked emitter (\hbox{$M_i=-26.8$}, Table~\ref{tabS}) and one of only three high-redshift ($z\sim0.6$) double-peaked emitters discovered serendipitously based on their H$\beta$-line profiles. Its inclusion in the current sample allows us to extend substantially the redshift and luminosity coverage of the sample. 

A simple power-law model modified by Galactic absorption in the \hbox{0.3--10\,keV} band results in a rather flat power-law slope of $\Gamma=1.0\pm0.1$, with large residuals around \hbox{$3.9-4.3$\,keV}, corresponding to \hbox{$\sim6.3-7$\,keV} in the AGN rest frame ($C$-stat=123 for 118\,$DoF$; see Figure~\ref{Xspec}).  The inclusion of neutral intrinsic absorption does not improve the fit. The addition of a Gaussian emission line centered at 6.7\,keV (rest-frame) with a width of 0.3\,keV improves the fit  ($C$-stat=118 for 117\,$DoF$), but the small number of counts prevents us from reliably constraining the line parameters. The equivalent width of the Gaussian line is $\sim0.7$\,keV, and if confirmed, this line is consistent with \ion{Fe}{25} K$\alpha$ emission. Note that the putative FeK$\alpha$ line EW is too high relative to the observed hard-band X-ray luminosity and would require unusually high Fe abundance; typical AGNs with \hbox{2--10}\,keV luminosities similar to that of  SDSS\,J2125$-$0813 ($10^{45}$\,erg\,s$^{-1}$) are expected to have EW$\sim40$\,eV \citep{Bianchi}.

More complex models including a partial or ionized absorber or an exponentially cut-off power-law model reflected by ionized material in addition to a relativistic Fe line provide better fits to the data, but cannot be well constrained with only 155 counts. For example, a model including an ionized absorber ($\xi\approx200$\,erg\,s$^{-1}$\,cm) with a large column density ($N\si{H,i}\sim10^{23}$\,cm$^{-2}$) provides a statistically acceptable fit, suggesting the slow wind outflow model of \citet{jet} might provide extra illumination to the H$\alpha$ emission region. A longer observation of this source is necessary to characterize better its X-ray emitting region.

SDSS\,J2125$-$0813 was detected in the RASS in November 1990, with a \hbox{0.5--2\,keV} flux of $(8\pm1)\times10^{-13}$\,erg\,cm$^{-2}$\,s$^{-1}$ and a hardness ratio of 0.86, which corresponds to a power-law photon index of $\Gamma_{0.1-2}=1.5$ (assuming absorption equal to the Galactic value). For comparison, the best-fit (Figure~\ref{Xspec}) model \hbox{0.5--2\,keV} flux in the \emph{Chandra} observation was $1.3^{+0.5}_{-0.5}\times10^{-13}$\,erg\,cm$^{-2}$\,s$^{-1}$, a soft X-ray band decrease in flux by a factor of  $\sim6$ over 15.5 years. The observed change in the soft-band flux is significant and robust to changes in the power-law model (e.g., assuming intrinsic absorption) or the X-ray band (e.g., restricting the analysis to the \hbox{0.5--2\,keV} band) used to obtain the flux during the \emph{Chandra} epoch.

\begin{deluxetable*}{ccccccccc} 
\tablewidth{0pt} 
\tabletypesize{\small}
\tablecaption{X-ray and Optical Fluxes, Luminosities and Flux-ratios} 
\tablehead{\colhead{Object} &\colhead{$F\si{2--10\,keV}$} &\colhead{$F\si{0.5--2\,keV}$} &\colhead{$l\si{2\,keV}$} &\colhead{$l\si{2500\,\AA}$} & \colhead{$\alpha\si{ox}$} &\colhead{$\Delta\alpha\si{ox}$} &\colhead{$\sigma_{\alpha\si{ox}}$} &\colhead{$L\si{bol}$}\\ 
\colhead{(1)} &\colhead{(2)} &\colhead{(3)} &\colhead{(4)} &\colhead{(5)} &\colhead{(6)} &\colhead{(7)} &\colhead{(8)}  &\colhead{(9)}\\
\colhead{HHMM$\pm$DDMM} &\colhead{erg\,cm$^{-2}$\,s$^{-1}$} &\colhead{erg\,cm$^{-2}$\,s$^{-1}$} &\colhead{$\log$[erg\,s$^{-1}$\,Hz$^{-1}$]} &\colhead{$\log$[erg\,s$^{-1}$\,Hz$^{-1}$]} &\colhead{} &\colhead{} &\colhead{} &\colhead{erg\,s$^{-1}$}}
\startdata
J0132$-$0952 & $8.7^{+2.1}_{-2.0}\times10^{-13}$ &  $4.8^{+1.2}_{-1.0}\times10^{-13}$ & 26.24 & 29.01 & $-$1.06 & 0.28 & 0.17 & $1\times10^{45}$\\ 
J1004$+$4801 & $2.9^{+0.6}_{-0.4}\times10^{-13}$ & $1.4^{+0.2}_{-0.2}\times10^{-13}$ & 25.45 & 28.22 & $-$1.06  & 0.17 & 0.19 & $2\times10^{44}$\\ 
J1238$+$5325 & $1.9^{+0.2}_{-0.2}\times10^{-12}$ & $12^{+1.2}_{-1.2}\times10^{-13}$ & 26.91 & 29.53 & $-$1.00 & 0.41 & ... & $4\times10^{45}$\\ 
J1407$+$0235 & $1.7^{+0.1}_{-0.1}\times10^{-12}$ & $8.1^{+0.6}_{-0.6}\times10^{-13}$ & 26.66 & 29.30 & $-$1.01 & 0.37 & 0.17 & $2\times10^{45}$\\ 
J2125$-$0813 & $7.1^{+2.1}_{-2.0}\times10^{-13}$ & $1.3^{+0.5}_{-0.6}\times10^{-13}$ & 26.77 & 30.23 & $-$1.33 & 0.18 & 0.20 & $1\times10^{46}$\\ 
\enddata
\tablecomments{(1) Short object name; (2) \hbox{2--10\,keV} rest-frame absorption-corrected flux; (3) \hbox{\hbox{0.5--2\,keV}} rest-frame absorption-corrected flux; (4) Logarithm of the 2\,keV monochromatic luminosity; (5)  Logarithm of the 2500\,\AA\ monochromatic luminosity; (6) UV-to-X-ray spectral index, $\alpha\si{ox}=-0.3838(l\si{2500\,\AA}-l\si{2\,keV})$; (7) $\alpha\si{ox}$ residuals, obtained by subtracting the $\alpha\si{ox}$ expected from the \hbox{$\alpha\si{ox}$($l\si{2500\,\AA}$)} relation of \citet{St06} from the observed one; (8) The spread of  similar luminosity RQ AGNs around the best-fit relation of \citet{St06}; (9) Bolometric luminosity estimate, see \S~\ref{bollum}.}
\label{tab3}
\end{deluxetable*}

\begin{deluxetable*}{cccccccccccc}
\tablewidth{0pt} 
\tablecaption{HET Observations and Balmer-Line Disk Fit Parameters}
\tablehead{\colhead{Object} &\colhead{ObsDate} &\colhead{$t\si{exp}$} &\colhead{$i$} &\colhead{$q$} &\colhead{$R_1$} &\colhead{$R_2$} &\colhead{$\sigma$ }&\colhead{$e$}  &\colhead{$\phi$}  &\colhead{$L\si{H$\alpha$}$} &\colhead{$L\si{H$\alpha$}/W\si{d}$} \\
\colhead{(1)} &\colhead{(2)} &\colhead{(3)} &\colhead{(4)} &\colhead{(5)} &\colhead{(6)} &\colhead{(7)} &\colhead{(8)} &\colhead{(9)}&\colhead{(10)} &\colhead{(11)} &\colhead{(12)}\\
\colhead{HHMM$\pm$DDMM} &\colhead{yyyy-mm-dd} &\colhead{sec} &\colhead{degree} &\colhead{} &\colhead{$R\si{G}$} &\colhead{$R\si{G}$} &\colhead{km\,s$^{-1}$}&\colhead{} &\colhead{degree}&\colhead{erg\,s$^{-1}$}&\colhead{}}
\startdata
J0132$-$0952 & 2006-08-23 & 1800 & 32 & 2 & 340 &   850 & 2400 & 0.1 & 50 & $5.6\times10^{42}$ & 0.4\\ 
J1004$+$4801 & 2006-03-29 & 2500 & 36 & 3 & 750 & 1470 & 2200 & $\cdots$  & $\cdots$  &  $1.4\times10^{42}$ & 1.8 \\   
J1238$+$5325 & 2006-06-17 & 900 & 48 & 2 & 400 & 1400 & 1800 & 0.3 & 30 & $15\times10^{42}$ & 0.3 \\ 
J1407$+$0235 & 2006-03-05 & 3200 & 20 & 2 & 180 &   210 & 1400 & 0.2 & 58 & $7.7\times10^{42}$ & 0.4 \\ 
J2125$-$0813 & 2006-06-23 & 1100 & 15 & 2 & 150 &   340 & 2000 & $\cdots$  & $\cdots$ & $9.1\times10^{42}$ & 0.1\\       
\enddata
\tablecomments{Accretion-disk model fits following \citet{CH89b} and \citet{E94}. (1) Short object name; (2) Date of the HET observation; (3) Exposure time, $t_{\textrm{exp}}$; (4) Disk inclination with respect of our line of sight; (5) Power-law index of the external illumination, $q$; the inner (6) and outer (7) radii of the emission region; (8) Turbulent broadening parameter; (9) Disk ellipticity; (10) Orientation angle of the disk semi-major axis with respect to our line of sight; (11) Luminosity of the H$\alpha$ line (or H$\beta$, in the case of SDSS\,J2125$-$0813); the uncertainties are on the order of \hbox{5\%--10\%}; (12) The ratio of the H$\alpha$-line luminosity to the viscous power available locally in the emission-line region of the accretion disk, $L\si{H$\alpha$}/W\si{d}$. The value quoted for SDSS\,J2125$-$0813 assumes a typical H$\alpha$-to-H$\beta$ flux ratio of 3.2.}
\label{tabFits}
\end{deluxetable*}

\begin{deluxetable*}{cccccccc}
\tablewidth{0pt} 
\tablecaption{Optical Variability: Time Scales and Amplitudes}
\tablehead{\colhead{Object} &\colhead{SDSS ObsDate} &\colhead{Galaxy} &\colhead{$\Delta t\si{HET-SDSS}$} &\colhead{$\Delta t\si{HET-\emph{Chandra}}$} &\colhead{$\Delta\lambda\si{rest}$} &\colhead{$F\si{SDSS}/F\si{HET}$} &\colhead{$F\si{H$\alpha$,SDSS}/F\si{H$\alpha$,HET}$}\\
\colhead{(1)} &\colhead{(2)} &\colhead{(3)} &\colhead{(4)} &\colhead{(5)} &\colhead{(6)} &\colhead{(7)} &\colhead{(8)} \\
\colhead{HHMM$\pm$DDMM} &\colhead{yyyy-mm-dd} &\colhead{\%} &\colhead{years} &\colhead{months} &\colhead{\AA} &\colhead{} &\colhead{}}
\startdata
J0132$-$0952 & 2001-08-26  & 20         & 5    & $-10$     & 4300--4600 & 1.15   & 0.9 \\ 
...                         & 2001-09-26  & 20         & 5    & $-10$     & 4300--4600 & 0.95   & 0.85 \\ 
J1004$+$4801 & 2002-03-06 & 70          & 4    & $1.9$    & 4500--7200 & 0.7  & 0.7 \\   
J1238$+$5325 & 2002-04-15 & $<$10   & 4.2 & $0.57$ & 4000--6200 & 0.53  & 0.63 \\ 
J1407$+$0235 & 2001-03-25 & 50          & 5    &  $4$      & 4000--4500 & 3        & 2.4  \\ 
J2125$-$0813 & 2001-09-01  & $<$10   & 4.7 & $3 $      & 3300--4200 & 0.45  & 0.4 \\       
\enddata
\tablecomments{(1) Short object name; (2) Date of the SDSS observation; (3) The percentage contribution from the host galaxy at 4200\,\AA\ from the SDSS spectrum, except for SDSS\,J1004$+$4801, for which the galaxy fraction is measured at 4500\,\AA; (4) Observed-frame time interval between the SDSS and the HET observations, $t\si{HET}-t\si{SDSS}$; the HET observation always followed the SDSS one; (5) Observed-frame time interval between the HET and the \emph{Chandra} observations, $t\si{HET}-t\si{\emph{Chandra}}$; a negative sign indicates that the HET observation precedes the \emph{Chandra} one; (6) Rest-wavelength range for which the flux change was measured; (7) The ratio of the HET to the SDSS flux; (8) The ratio of the HET to the SDSS H$\alpha$-line (H$\beta$ in the case of SDSS\,J2125$-$0813) flux.}
\label{tabVar}
\end{deluxetable*}

\begin{figure}
\plotone{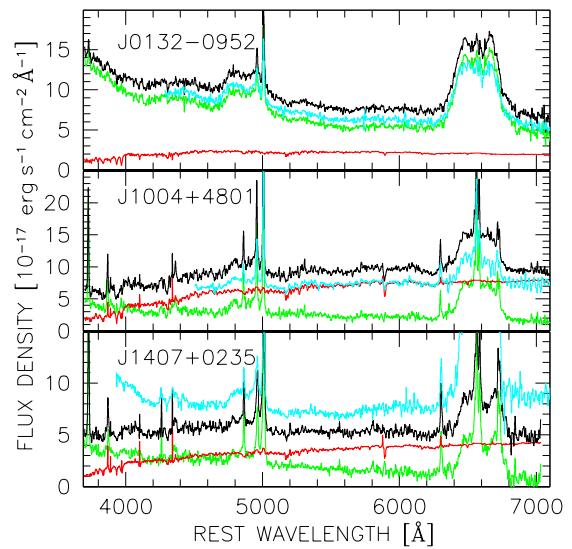}
\caption{AGN-host-galaxy decomposition for the three double-peaked emitters whose SDSS spectra have detectable host-galaxy contributions.  The host galaxy component is shown in red, the AGN component in green, the total SDSS spectrum in black, and the total HET spectrum in cyan. All spectra are observed flux-calibrated spectra, i.e., no Galactic extinction correction has been applied. The SDSS spectra are convolved with an 8\,\AA\ Gaussian kernel for display purposes. The top panel shows only one of the two SDSS spectra, taken on 2001 September 26.
\label{OspecFit}}
\end{figure}

\section{Contemporaneous Optical Observations}
\label{optspec}

We re-observed all five double-peaked emitters with the Marcario Low-Resolution Spectrograph (LRS) of the Hobby-Eberly Telescope  (HET) in the optical band, within a few months of the \emph{Chandra} observations. Double-peaked emitters are known to vary on timescales of months to years \citep[e.g.,][]{var1,var2},  and thus contemporaneous observations are necessary to ensure proper comparison of the optical and the X-ray monochromatic luminosities. In addition to flux variations, the line profile shapes also vary on timescales of months to years and a contemporaneous spectrum allows us to estimate the emission-region size during the \emph{Chandra} epoch. We used the G1 grism and 1$''$ slit to cover the wavelength range \hbox{4150--10100\,\AA}.  Our spectra have a spectral resolution of 8.4\,\AA, corresponding to a velocity resolution of 350\,km\,s$^{-1}$ at 7200\,\AA. The data were reduced in a standard manner, as described in \citet{E94}, for example.

We used the [\ion{O}{3}] line flux measured from the SDSS spectra to determine the flux calibration of the HET spectra. The [\ion{O}{3}] line emission is a good tracer of nuclear activity in AGNs  \citep[e.g.][ and references therein]{K03} and provides a suitable and relatively robust measure (we estimate $\lesssim10$\% uncertainty) of the spectral normalization. 

We isolated the H$\alpha$-line profiles (or the H$\beta$ profile in the case of SDSS~J2125$-$0813) in the HET/LRS spectra by subtracting a local power-law continuum. We fitted the resulting line profiles with the relativistic Keplerian disk models of \citet{CH89b} and \citet{E95} in order to determine the parameters that characterize the line-emitting region of the disk. For these fits we ignore the regions around the narrow H$\alpha$ line as well as the S, N, and O forbidden lines. We used the SDSS line-profile fits, which include modeling of  the narrow lines, to verify that the results obtained from the HET spectra are robust. The line-profile fits allow us to estimate the following parameters: (1) the inner and outer radii of the emission region, $R_1$ and $R_2$, in units of $R_G\equiv GM/c^2$, where $M$ is the mass of the black hole, (2) the inclination of the axis of the disk to the line of sight, $i$, (3) the emissivity power-law index, $q$,\footnote{The external illumination is usually assumed to be well described by a power law with an index $q$ with $2\leq q \leq3$, where $q=3$ represents a point source high above the disk on the disk axis, while $q=2$ represents a constant-density illuminator above the emission region \citep[see][]{dumont90}.} under the convention that $\epsilon \propto R^{-q}$, and (4) the broadening parameter, $\sigma$ (in km~s$^{-1}$). 

The line profiles of the five objects of our sample are shown in Figure~\ref{Ospec}, together with the best-fitting disk models. Table~\ref{tabFits} lists the model parameters. Two of the objects are well fit by an axisymmetric disk model, and all disk parameter estimates are fairly robust (in the case of SDSS\,J2125$-$0813, for example, the inner radius of the emission region is constrained to lie between $110\,R\si{G}<R_1<180\,R\si{G}$). The remaining three objects require non-axisymmetric disk models, and we model them with an elliptical disk. The elliptical disk model \citep{E95} is the simplest non-axisymmetric accretion-disk model which we use to obtain estimates of the inner and outer radii of the H$\alpha$-line emission region; besides the five disk parameters listed above, this model requires fitting for the disk ellipticity, $e$, and the angle of the semi-major axis with respect to our line of sight, $\phi$. The inner and outer radii of the emitting region are less well constrained in these three cases, but choosing models with fixed emissivity power-law indices,  we can estimate the inner radius to about $\pm$30\%. We start by assuming either a point-source illumination ($q=3$) or illumination by scattering ($q=2$), whenever the point-source model fit is inadequate. We were able to constrain the inclination angles to within a few degrees in all cases, which is important since this parameter is degenerate with the inner radius. The inner radius of the emitting region is a crucial parameter for the determination of the viscous energy available locally. The ratio of the line luminosity to the locally available energy (listed in Table~\ref{tabFits} in all cases)  detemines the need for external illumination as discussed in \S~\ref{diskP} below.

We use the bluest parts of the HET spectra (3300\,\AA\ for SDSS\,J2125$-$0813,  and \hbox{4000--4500\,\AA} for the remaining targets) to estimate the change in flux between the HET and SDSS epochs and list the results in Table~\ref{tabVar}. Table~\ref{tabVar} summarizes the optical variability time scales and amplitudes for the five \emph{Chandra} double-peaked emitters. The changes in the continuum fluxes over the wavelength ranges specified in column 6 are given in column 7, with the changes in the H$\alpha$-line fluxes (H$\beta$ in the case of SDSS\,J2125$-$0813) listed in column 8. The continuum flux varied by factors of 2--3 for the three higher redshift sources (SDSS\,J1238$+$5325, SDSS\,J1407$-$0235, and SDSS\,J2125$-$0813) in the 3--4\,year (rest-frame) period between the SDSS and HET observations. In addition to the reported continuum-flux variability, the line profile shapes also varied between the SDSS and HET observations. The ratio of the blue-to-red peak height varied from $\sim$0.9 to $\sim$1 in the case of SDSS\,J0132$-$0952, and $\sim1.1$ to $\sim0.8$ in the case of SDSS\,J1407$-$0235 (see the last panel of Figure~\ref{Ospec}), as has been observed for other double-peaked emitters \citep[e.g.,][]{var1,var2}.  

In order to estimate the host-galaxy corrections, we fit the SDSS spectra, which have broader wavelength coverage and provide better constraints on the AGN--host-galaxy decomposition, with a combination of host-galaxy and AGN components. The results of these fits, for the three double-peaked emitters with detectable host-galaxy contribution in the 3$''$-diameter SDSS fibers, are shown in Figure~\ref{OspecFit}. The  host-galaxy and AGN components for the fit are constructed using AGN and galaxy eigenspectra obtained from large SDSS samples by \citet{Yip1,Yip2}. We subtract the host-galaxy spectra obtained in this fashion from the three SDSS spectra with substantial host-galaxy contributions: SDSS\,J0132$-$0952, SDSS\,J1004$+$4801, and SDSS\,J1407$-$0235 (see Table~\ref{tabVar} for the respective host-galaxy contributions).

In order to extrapolate the optical observations to the UV and obtain an estimate of the monochromatic luminosities at 2500\,\AA, $l\si{2500\,\AA}=\log(L\si{2500\,\AA})$, we scale the mean quasar spectral energy distribution (SED) of \citet{Richards06} to match the observed AGN continua in the optical. We use the HET spectra of SDSS\,J1238$+$5325 and SDSS\,J2125$-$0813 for this purpose, and the SDSS AGN components scaled to the HET epoch for the three sources with significant galaxy contributions. This is necessary as the HET coverage is insufficient to provide good constraints on the host-galaxy contributions. We caution, however, that this solution is not optimal, since the 3$''$-diameter SDSS fibers are larger than the rectangular $1'' \times 10''$ LRS slits, and the host-galaxy contribution could be smaller in the HET spectra. Under the seeing conditions at which the HET spectra were obtained, however, the AGN spectrum is further diluted by starlight and this representation is adequate. The \citet{Richards06} quasar SED was constructed to cover the infrared (IR) to the X-ray region using \emph{Spitzer Space Telescope}, SDSS, GALEX (GALaxy evolution EXplorer), and \emph{ROSAT} data for a sample of unobscured SDSS quasars. The slopes of the host-galaxy corrected spectra for all of our sources agree well with the \citet{Richards06} quasar continuum, as discussed further in \S~\ref{bollum}.

\begin{deluxetable*}{cccccc}
\tablewidth{0pt} 
\tablecaption{GALEX UV Observations}
\tablehead{\colhead{Object} &\colhead{GALEX ObsDate} &\colhead{$F_{NUV}$} &\colhead{$F_{FUV}$} &\colhead{$E(B-V)$} &\colhead{$A_{FUV}$} \\
\colhead{(1)} &\colhead{(2)} &\colhead{(3)} &\colhead{(4)} &\colhead{(5)} &\colhead{(6)}  \\
\colhead{HHMM$\pm$DDMM} &\colhead{yyyy-mm-dd} &\colhead{$\mu$Jy} &\colhead{$\mu$Jy} &\colhead{} &\colhead{mag} }
\startdata
J0132$-$0952 & 2007-10-22  & $55.2\pm3.9$   & $31.1\pm4.0$   & 0.028  & 0.22  \\
J1004$+$4801& 2007-02-10  & $16.3\pm3.1$   & $10.3\pm3.0$   & 0.0092 & 0.073 \\
J1238$+$5325 & 2005-03-02 & $215.0\pm5.6$ & $156.5\pm8.9$ & 0.015  & 0.12   \\
J1407$+$0235 & 2004-05-17 & $57.2\pm1.4$   & $48.9\pm1.9$   & 0.037   & 0.29   \\
J2125$-$0813 & 2003-08-27  & $112.7\pm6.1$ & $27.0\pm5.8$   & 0.056   & 0.44 \\
\enddata
\tablecomments{(1) Short object name; (2) Date of the GALEX observations; (3) Observed near-UV ($NUV$) flux; (4) Observed far-UV ($FUV$) flux; (5) Galactic reddening following \citet{Schlegel}; (6) $FUV$ extinction correction, \hbox{$A_{FUV}=7.9E(B-V)$}, using the \citet{Cardelli} parameterization of the Galactic extinction law for total-to-selective extinction ratio of $R_V=3.1$; the extinction correction in the $NUV$, \hbox{$A_{NUV}=8.0E(B-V)$}, is the same within 2--3\%.}
\label{tabGalex}
\end{deluxetable*}

\begin{deluxetable*}{ccccc}
\tablewidth{0pt} 
\tablecaption{2MASS NIR Observations}
\tablehead{\colhead{Object} &\colhead{2MASS ObsDate} &\colhead{$J$} &\colhead{$H$} &\colhead{$K_s$} \\
\colhead{(1)} &\colhead{(2)} &\colhead{(3)} &\colhead{(4)} &\colhead{(5)}  \\
\colhead{HHMM$\pm$DDMM} &\colhead{yyyy-mm-dd} &\colhead{mag} &\colhead{mag} &\colhead{mag} }
\startdata
J0132$-$0952 & 1999-08-09  & $16.93\pm0.20$  & $16.00\pm0.20$ & $14.71\pm0.12$ \\ 
J1004$+$4801& 1999-01-19  & $16.32\pm0.11$  & $15.58\pm0.14$ & $14.97\pm0.13$ \\ 
J1238$+$5325 & 2000-04-25 & $16.04\pm0.11$ & $15.43\pm0.15$  & $14.55\pm0.10$  \\ 
J1407$+$0235 & 2000-03-28 & $>17$                   & $>16.3$                 & $>15.8$                \\ 
J2125$-$0813 & 2000-08-27  & $16.02\pm0.10$ & $15.16\pm0.11$  & $14.28\pm0.08$  \\ 
\enddata
\tablecomments{(1) Short object name; (2) Date of the 2MASS observation; (3) Observed $J$-band magnitude. The $J$ band has an effective wavelength of  $\lambda\si{J,eff}=1.235$\,$\mu$m; (4) Observed $H$-band magnitude, $\lambda\si{H,eff}=1.662$\,$\mu$m; (5) Observed $K_s$-band magnitude, $\lambda\si{K$_s$,eff}=2.159$\,$\mu$m. Note that the Galactic reddening corrections are $<0.05$\,mag in all cases for all three bands.}
\label{tab2MASS}
\end{deluxetable*}

\section{DISCUSSION AND CONCLUSIONS}
\label{conclusions}

\subsection{X-ray Continua}

Four of the five \emph{Chandra} double-peaked emitters are well fit by power-law models with spectral indices in the range $1.6<\Gamma<1.8$, in good agreement with the continua of similar luminosity AGNs, $\left< \Gamma \right> = 1.74\pm0.09$, studied by \citet{piconcelli}. A well-known correlation exists between $\Gamma$ and the FWHM the H$\beta$ line in AGNs \citep[e.g.,][]{Brandt97}. \citet{Shemmer06,Shemmer08} show that the $\Gamma$--FWHM$\si{H$\beta$}$ relation is driven by the relation between the hard X-ray band photon index and the accretion rate (parameterized by $L/L\si{Edd}$), since FWHM$\si{H$\beta$}$ is an accretion-rate indicator for unobscured AGNs. There is no evidence that the Balmer-line widths can be used as an accretion-rate indicator for double-peaked emitters, especially for the double-peaked emitters with extremely large line widths considered here. However, if the X-ray power-law photon index can be used as an indicator of the accretion rate, the double-peaked emitters studied here would have $L/L\si{Edd}\approx 0.1$ (with a $1\sigma$ range of $0.005<L/L\si{Edd}<1$) and at least four of the five would be considered efficient accretors, a result that is supported further by the existence of ``big blue bumps" -- a characteristic feature in the spectral energy distributions of quasars (see \S~\ref{bollum}).
                   
\subsection{X-ray and Optical Variability}

Four of the five double-peaked emitters observed with \emph{Chandra} are X-ray variable on long timescales. Both SDSS\,J0132$-$0952 and SDSS\,J1407$+$0235 have RASS upper limits lower than the observed ACIS-S fluxes in the \hbox{0.5--2\,keV} band, suggesting that these sources have brightened on decade timescales by at least a factor of $\gtrsim1.6$ and $\gtrsim4$, respectively. SDSS\,J1238$+$5325 has the same X-ray flux during the two observations separated by 12 years in the AGN rest-frame. The remaining two sources are observed in a fainter state with \emph{Chandra} than their corresponding early 1990s \emph{ROSAT} PSPC  detections by factors of  about 3 (SDSS\,J1004$+$4801) and 6 (SDSS\,J2125$-$0813). For comparison, \citet{Shinozaki} find that the typical flux-variability amplitude of similar luminosity AGNs on timescales of decades is $\sim2.5$. \citet{Mateos} compute the mean flux-variability amplitude, $\sigma\si{intr}$ \citep[for a description of this quantity see][]{Almani}, of the unobscured AGNs from the Lockman Hole sample. They find that the average AGN has a $\sigma\si{intr}\sim0.3$, with a range of $0.2<\sigma\si{intr}<0.5$ for the AGNs with confirmed variability. The corresponding $\sigma\si{intr}$ values for our four variable sources are $>$0.2, 0.5, $>$0.6, and 0.7 for SDSS\,J0132$-$0952, SDSS\,J1004$+$4801, SDSS\,J1407$-$0235, and SDSS\,J2125$-$0813, respectively. Based on this small sample, it appears that most double-peaked emitters are more variable on decade timescales than the typical AGN. 

We tested the \emph{Chandra} light curves for short-term variability using the XRONOS package. It enables us to compute the RMS fractional variation  ($\sigma\si{RMS}$) of the count rate within each exposure after binning the data in intervals ranging from half a minute to about 15--30 minutes (allowing for at least 5 bins in each case). No short-term variability was detected. The $3\sigma$ Gaussian limits on the variability range between $\sigma\si{RMS}<0.1$--0.3 for the higher signal-to-noise low-redshift objects and $\sigma\si{RMS}<0.4$--0.6 for SDSS\,J2125$-$0813, depending on the length of the time interval probed.  The non-detection of short-term variability is expected, as our objects are relatively luminous in the \hbox{0.5--10\,keV} band (with luminosities ranging between a few times $10^{43}$\,erg\,s$^{-1}$ and $10^{45}$\,erg\,s$^{-1}$), and AGN variability amplitudes on short timescales are known to decrease with increasing luminosities \citep[e.g.,][]{Nandra97}. \citet{LP93} show that for AGNs with luminosities in the $10^{43}-10^{45}$\,erg\,s$^{-1}$ range, timescales longer than $10^4-10^6$\,s are typically necessary to detect RMS variability of more than 10\%; these timescales are 2--10 times longer than the exposure times in our snapshot survey.

The optical variability of the five double-peaked emitters is also significant on long timescales (see Table~\ref{tabVar}), with flux-density changes of up to a factor of 3 over 3--4\,yrs in the AGN rest frame in our sample, accompanied by H$\alpha$-line flux changes of up to a factor of 2.5. Figure~\ref{OspecFit} shows the HET and SDSS spectra of three of the five \emph{Chandra} sources. Only two of the double-peaked emitters (SDSS\,J0132$-$0952 and SDSS\,J1407$+$0235) show clear H$\alpha$ line-profile variations which are similar to those observed in double-peaked emitters with long-term variability monitoring \citep[e.g.,][]{var1,var2}. The three higher  redshift AGNs in our sample (SDSS\,J1238$+$5325, SDSS\,J1407$+$0235, and SDSS\,J2125$-$0813) vary by factors of \hbox{$\sim$2--3} in the 4000\,\AA\ continuum in a few years. For comparison, the expected AGN variability at 4000\,\AA\ for a similar time lag in the SDSS is $\sim0.2$\,mag, or about 60\% \citep{sdssvar}. 

We conclude that the double-peaked emitters studied here are typically more variable on long timescales (years) in the optical (a variability range of 10--300\%, see Table~\ref{tabVar}) and the X-ray (factors of $\gtrsim1.6$, 3, 1, $\gtrsim4$, and 6, in order of increasing RA) than the average AGN. In S06 we compared the  width of the $\alpha\si{ox}$ residual distribution to that of the luminosity-matched \citet{St06} subsample and found that it was broader at the 1$\sigma$ level. This finding, which could be interpreted as an indication of larger X-ray and/or optical/UV variability for double-peaked emitters, is confirmed with the small sample studied here. 

\begin{figure}
\plotone{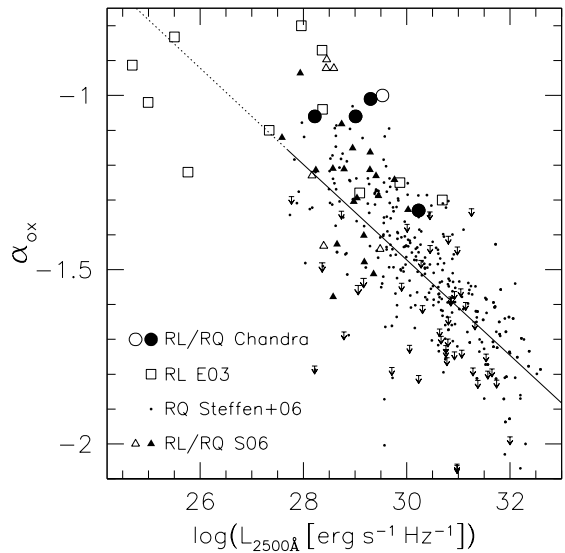}
\caption{The UV-to-X-ray spectral index vs.\ the 2500\,\AA\ monochromatic luminosity. The solid line is the \citet{St06} $\alpha\si{ox}$-$l\si{2500\,\AA}$ relation; the dotted part is an extrapolation. 
\label{aox}}
\end{figure}

\subsection{Distribution of the UV-to-X-ray Indices}
\label{Saox}

We use the UV-to-X-ray index, $\alpha\si{ox}$, defined in terms of the logarithm of the ratio of the rest-frame 2500\,\AA\ and 2\,keV monochromatic fluxes, $\alpha\si{ox}=-0.3838\log[F_{\nu}(\textrm{2500\,\AA})/F_{\nu}(\textrm{2\,keV})]$, to compare the X-ray emission of our sample to that of more typical AGNs with similar optical/UV monochromatic luminosities. The UV-to-X-ray indices of the five \emph{Chandra} targets are shown as a function of their 2500\,\AA\ monochromatic luminosities in Figure~\ref{aox} and listed in Table~\ref{tab3}. There is a tendency for the double-peaked emitters to have flatter $\alpha\si{ox}$ values than typical RQ AGNs. This is expected for the one RL double-peaked emitter, SDSS\,J1238$-$5325, but is also apparent in the four RQ targets. We can characterize this discrepancy by subtracting from each observed value of $\alpha\si{ox}$ the value expected based on the $\alpha\si{ox}(l\si{2500\,\AA})$ relation found by \citet{St06} for RQ AGNs, $\alpha\si{ox}(l\si{2500\,\AA})=-0.137l\si{2500\,\AA}+2.638$, and compare this to the spread, $\sigma\si{$\alpha\si{ox}$}$, observed around the relation for normal RQ AGNs. The $\alpha\si{ox}$ residual values ($\Delta\alpha\si{ox}$), together with the values of $\sigma\si{$\alpha\si{ox}$}$ for similar luminosity RQ AGNs are given in columns 7 and 8 of Table~\ref{tab3}.  

The $\alpha\si{ox}$ residuals of the four RQ double-peaked emitters are larger than or about equal to the spread observed around the $\alpha\si{ox}(l\si{2500\,\AA})$ relation for normal RQ AGNs. The $\alpha\si{ox}$ values are $\sim 2 \sigma\si{$\alpha\si{ox}$}$ too flat in the cases of SDSS\,J0132$-$0952 and SDSS\,J1407$+$0235, and $\sim 1 \sigma\si{$\alpha\si{ox}$}$ too flat in the cases of SDSS\,J1004$+$4801 and SDSS\,J2125$-$0813.  The remaining double-peaked emitter, SDSS\,J1238$+$5325, has  an UV-to-X-ray index which is 0.41\,dex flatter than the value expected for similar-luminosity RQ AGN. This difference is typical for RL AGNs, which are known to have 2\,keV monochromatic luminosities of a factor of $\gtrsim3$ higher than those of RQ AGNs with similar UV emission \citep[e.g.,][]{rlagn}. The combined optical and X-ray variability of the double-peaked emitters together with the uncertainties in the determination of $l\si{2500\,\AA}$ is large enough to account at least partially for the $\Delta\alpha\si{ox}$ values observed for SDSS\,J1004$+$4801 (up to 0.2\,dex), SDSS\,J1407$+$0235 (up to 0.3\,dex), and SDSS\,J2125$-$0813 (up to 0.3\,dex). For X-ray or optical variability to be the cause of the anomalous $\alpha\si{ox}$ values observed for these three RQ AGNs, the variability amplitude over the 2--4 months separating the \emph{Chandra} and HET observations would have to reach the amplitude observed over years (optical) or decades (X-ray). Since the amplitude of AGN variability typically increases with time, this is not likely.

Using alternative methods to estimate $l\si{2500\,\AA}$, by (1) extrapolating the measured flux density at 3700\AA\ to 2500\,\AA\  using a uniform power-law of the form, $f_{\lambda} \propto \lambda^{\alpha_{\lambda}}$,  with slope $\alpha_{\lambda}=-1.5$ \citep{Berk}, or (2) using the \citet{Richards06} SED template scaled to the observed fluxes in the extreme UV (see \S~\ref{bollum}), we find results consistent with those reported in Table~\ref{tab3} within 0.2\,dex (typically $<$0.1\,dex)  in all cases. This is equivalent to an uncertainty of $<$0.1\,dex in $\alpha\si{ox}$, which is inadequate to explain the observed flattening of the UV-to-X-ray index.

In addition, although variability and measurement uncertainty could be responsible for the flattening of the $\alpha\si{ox}$ values in individual cases, it is unlikely that all four RQ double-peaked emitters would be affected in the same way. The mean value of the $\alpha\si{ox}$ residuals is  $\left<\alpha\si{ox}-\alpha\si{ox}(l\si{2500\,\AA})\right>=0.25\pm0.04$ for the four RQ double-peaked emitters observed with \emph{Chandra}, compared to  $\left<\alpha\si{ox}-\alpha\si{ox}(l\si{2500\,\AA})\right>=-0.02\pm0.01$, for the luminosity-matched subsample of RQ AGNs from \citet{St06}. Both the Gehan and logrank tests ($T=3.2$, $P=0.1$\% and $T=7.7$, $P<0.1$\%, respectively; see S06 for the use of these sample-comparison tests) confirm that the current sample of 4 RQ double-peaked emitters has a statistically different distribution of $\alpha\si{ox}$ residuals from the one found for all AGNs with similar UV luminosities and single-peaked lines presented in \citet{St06}. This tendency was discernible in the S06 sample (left panel of Figure~9 in S06), but not statistically significant; it was also observed in the preliminary analysis of the five \emph{Chandra} targets in combination with a sample of  double-peaked emitters observed with \emph{Swift} \citep{S07}. 

\subsection{Bolometric Luminosities}
\label{bollum}

\begin{figure}
\plotone{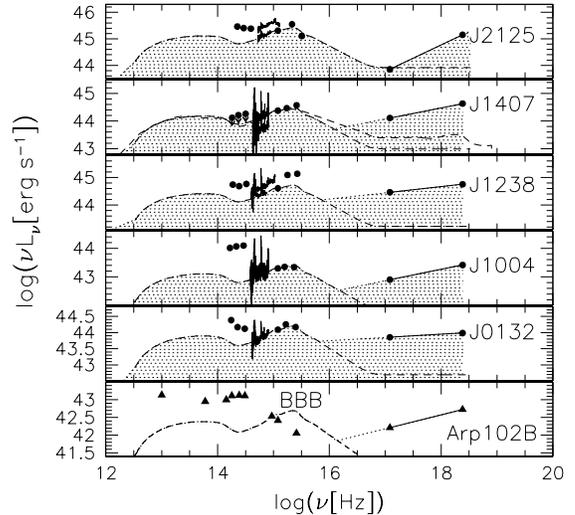}
\caption{NIR through X-ray SEDs of the five double-peaked emitters from the \emph{Chandra} sample in comparison with the SED of Arp\,102B (see \S~\ref{bollum} for details). The observations are shown with solid lines and points. The \citet{Richards06} SED templates, normalized to the optical continua during the HET epoch, are shown with dashed lines in each panel. The SDSS\,J1407$-$0235 panel also shows the RQ SED of \citet{Elvis}, with a dashed line that is distinctly higher in the X-ray range compared to the \citet{Richards06} SED. An SED template was created in each case by extrapolating the observed X-ray power law into the extreme UV, until it meets the optically-normalized \citet{Richards06} template; the area under the SED template, used to compute the bolometric luminosities, is shaded in each panel. The bottom panel shows the Arp\,102B SED in comparison to the \citet{Richards06} SED, normalized at 3200\,\AA; the Arp\,102B SED lacks a Big Blue Bump (BBB) around $\nu\sim10^{15.4}$\,Hz.
\label{sed}}
\end{figure}

Figure~\ref{sed} shows the spectral energy distributions (SEDs) of the five \emph{Chandra} sources from the near-infrared (NIR) to the X-ray band. The X-ray spectra are represented as power laws with photon indices given in Table~\ref{tab2}. For the two AGN with negligible host-galaxy contamination (SDSS\,J1238$+$5325 and SDSS\,J2125$-$0813) we show the HET-epoch spectra in the optical region; for the remaining three objects we plot the SDSS AGN components scaled to the flux level during the HET epoch. In all five cases  we also plot the SDSS-epoch AGN components. Comparison of the HET-epoch and SDSS-epoch spectra gives an idea of the optical variability on long timescales (4--5\,years). The near-UV (NUV, $\lambda\si{eff}=2267$\,\AA) and far-UV (FUV, $\lambda\si{eff}=1516$\,\AA) fluxes, obtained by the GALEX mission (see Table~\ref{tabGalex}) and corrected for Galactic extinction, are also shown. The extinction corrections, $A_{FUV}=7.9E(B-V)$ and $A_{NUV}=8.0E(B-V)$, are from \citet{dePaz}, who used the  \citet{Cardelli} parameterization of the Galactic extinction law. The NIR fluxes (obtained from the 2\,MASS magnitudes reported in Table~\ref{tab2MASS}) contain the host-galaxy contributions in all cases, and are well above the template SEDs in Figure~\ref{sed}. If we subtract an elliptical-galaxy SED (scaled to the galaxy contribution in the SDSS spectra and corrected for the SDSS fiber apertures; see Fioc \& Rocca-Volmerange 1997 for the galaxy SED) from the NIR fluxes, we can account for part of the discrepancy between the observed NIR fluxes and the SED templates for SDSS\,J0132$-$0952 and SDSS\,J1004$+$4801. In both cases a standard elliptical-galaxy template can account for $\lesssim$0.1\,dex in the NIR; a dusty starburst would be necessary to explain the high observed NIR luminosities. In the cases of SDSS\,J1238$+$5325 and SDSS\,J2125$-$0813, the NIR colors are more consistent with an AGN SED and the high observed luminosities are in line with the higher optical/UV fluxes.

The average quasar SED from \citet{Richards06}, scaled to the AGN continuum flux in the optical in each case, is also shown in Figure~\ref{sed}. We used the SDSS AGN component, scaled to the HET-epoch flux in the three cases with significant host-galaxy contribution, and the HET spectra in the remaining two cases. The \citet{Richards06} average SEDs are similar from the infrared (IR) to the UV region to the standard \citet{Elvis} median SEDs (both RQ SED templates are shown in the second panel of Figure~\ref{sed} for one of our sources, SDSS\,J1407$+$0235), but they differ substantially in the X-ray band. The \citet{Elvis} sample was biased toward AGNs with good S/N \emph{Einstein} data (hence X-ray brighter than the typical AGN) and bright in the extreme UV, and \citet{Richards06} argue that typical AGNs are redder and have lower X-ray-to-optical ratios. The \citet{Richards06} RQ SED has an $\alpha\si{ox}=-1.53$ and the \citet{Elvis} RQ SED an $\alpha\si{ox}=-1.38$, compared to $\alpha\si{ox}=-1.55$ and  $\alpha\si{ox}=-1.54$ for their respective $l\si{2500\,\AA}$ from the \citet{St06} relation; both SEDs have total (including the IR bump) bolometric luminosities of $2\times10^{46}$\,erg\,s$^{-1}$. Note that the GALEX observations agree well with the optically-scaled \citet{Richards06} template SEDs for the four RQ \emph{Chandra} AGN (within 0.11\,dex, or $<30$\%).  The \emph{NUV} band and the optical bracket rest-frame 2500\,\AA\ in all four cases, suggesting that the $l\si{2500\,\AA}$ estimates based on the optically-scaled teplate SEDs and the UV-to-X-ray indices are fairly robust (changes in $\alpha\si{ox}$ of less than 0.05). The RL AGN, SDSS\,J1238$+$5325, is 0.4\,dex more luminous in the \emph{NUV} band during the GALEX epoch relative to the HET-epoch optically-scaled RL template SED; a similar variation at 2500\,\AA\ in the 17 days (13 in the AGN rest frame) between the \emph{Chandra} and HET observations would result in a change in $\alpha\si{ox}$ of 0.16, and cannot fully account for the observed flatter $\alpha\si{ox}$.

The Arp\,102B SED, discussed in more detail by \citet{ArpASCA}, is shown in the last panel of Figure~\ref{sed} for comparison.  We use the \emph{Advanced Satellite for Cosmology and Astrophysics} (\emph{ASCA}) X-ray spectrum presented by Eracleous et al.~(2003a; see also Lewis \& Eracleous 2006), the \emph{Hubble Space Telescope} (\emph{HST}) UV spectrum given in \citet{Halpern96}, and the NIR and  mid-infrared (MIR) spectra published by \citet{Riffel} and \citet{ArpSpitzer}, respectively. \citet{Riffel} note strong stellar features in the NIR spectrum of Arp\,102B, which was integrated over a 700\,pc region centered on the active nucleus. The MIR source extent, studied by \citet{ArpSpitzer}, was around 250\,pc; for comparison, the \emph{HST} aperture diameter is $\sim$400\,pc. In the same panel of Figure~\ref{sed}, we show the average quasar SED from \citet{Richards06} in the infrared-to-UV region (scaled to the reddest UV wavelength of the Arp\,102B \emph{HST} spectrum, $\sim$3200\,\AA) together with the \emph{ASCA} power law in the X-ray region. In addition, the NIR emission of Arp\,102B includes a strong galaxy contribution, while in the MIR variability by a factor of 2.5 over a five year period was noted; both variability and varying host-galaxy contributions are likely responsible for part of the discrepancy between the quasar SED and the Arp\,102B IR data. The Arp\,102B SED is characterized by a lack of a `big blue bump' (BBB), the excess emission between the X-ray and UV bands ($\nu\sim10^{15.4}$\,Hz) normally associated with the innermost parts of the accretion disk (which is likely replaced by a hot radiatively inefficient central structure in the case of Arp\,102B and other AGNs with RIAFs). In contrast to the Arp\,102B SED, the GALEX observations of the five double-peaked emitters presented here show that the BBB is present in all five SEDs, with the possible exception of SDSS\,J1004$+$4801. The presence of BBBs is further confirmation for the relatively high efficiency of the accretion flows in these AGNs, excluding RIAF models for them.

Using the template SEDs from Figure~\ref{sed}, we can estimate the bolometric luminosities of the double-peaked emitters which we list in the last column of Table~\ref{tab3}. The \citet{Richards06} SED does not extend to the radio domain ($10^9\textrm{\,Hz}<\nu<10^{12.5}\textrm{\,Hz}$), but the $\nu<10^{12.5}$\,Hz region is energetically unimportant and does not affect the bolometric luminosity estimates. The template SED bolometric luminosities will overestimate the true luminosities if the IR bump consists of fully reprocessed nuclear continuum emission and should be excluded when computing the total luminosity.  Such `double-counting' of the IR emission can result in overestimates of the bolometric luminosity of up to 20--30\%. We can obtain upper limits to the bolometric luminosities and an estimate of possible uncertainties by normalizing the  IR-to-extreme-UV part of the SED template to the NIR (in all cases but SDSS\,J1004$+$4801), the extreme UV (in the case of SDSS\,J1238$+$5325), and/or the SDSS-epoch spectra (in the case of SDSS\,J2125$-$0813, for example, where the SDSS-epoch flux was higher). In all cases, we find that the bolometric luminosities can be at most a factor of 2--3 higher than those reported in Table~\ref{tab3}.

On average about 30\% of the bolometric luminosity is emitted between \hbox{0.5--10}\,keV in our sample (with a range of 16--36\%; or between 2\% and 12\% in the 0.5--2\,keV X-ray band), compared to $\sim3$\% emitted between \hbox{0.5--10}\,keV for the AGN composites of \citet{Richards06} and $\sim8$\% for the median RQ AGN composite of \citet{Elvis}. This comparison is not entirely fair, since the \citet{Richards06}  and \citet{Elvis} SEDs are constructed using more luminous AGNs ($l\si{2500\,\AA}=30.6$ and $l\si{2500\,\AA}=30.5$, respectively) and $\alpha\si{ox}$ is luminosity dependent. If we renormalize the \citet{Richards06} SED to the average value of  $l\si{2500\,\AA}$ in our sample, $l\si{2500\,\AA}=29.2$, and scale the X-ray part of the SEDs to the level inferred from the \citet{St06}  $\alpha\si{ox}(l\si{2500\,\AA})$ relation for typical RQ AGNs,  we find that even in this case only 7--8\% of the bolometric luminosity is emitted between \hbox{0.5--10}\,keV by typical RQ AGNs. This restates the conclusion from \S~\ref{Saox}: the average RQ  double-peaked emitter from our \emph{Chandra} sample has an $\alpha\si{ox}$ value flatter by 0.25\,dex relative to that of a typical RQ AGN and is 0.65\,dex more luminous at 2\,keV, which is equivalent to emitting about 30\% of its bolometric luminosity between \hbox{0.5--10}\,keV.

\subsection{Disk Illumination}
\label{diskP}

To compute theoretically the emerging spectrum of an accretion disk, both an accurate representation of the vertical structure of the disk and a suitable approximation for the radiation transfer are necessary. \citet{Williams80}, for example, computes the Balmer-line emission from the outer, optically thin parts of the accretion disks of cataclysmic variables, assuming a vertically homogeneous disk and local termodynamic equilibrium (LTE). This paper finds that the outer regions of the disk cool through line emission (with $<20$\% of the locally available viscous energy emitted as H$\alpha$), producing the Balmer-line series with small Balmer decrements ($F\si{H$\alpha$}/F\si{H$\beta$}\approx1$). For comparison, the observed Balmer-line decrements range between 3 and 4 for the four low-redshift \emph{Chandra} targets (for which both H$\alpha$ and H$\beta$ fall within the SDSS wavelength coverage) similar to the Balmer decrement of 3.2 which is considered typical for the BLR gas.

Assuming a homogeneous photoionized disk, \citet{CS87} and \citet{dumont90} argued that standard accretion disks in AGNs can produce low-ionization lines with characteristics similar to those observed in the BLRs of AGNs (including steeper Balmer-line decrements) if the disk is externally illuminated. The lines arise in the radiatively excited skin of the inner ($10^2 R\si{G}<R<10^4 R\si{G}$) accretion disk or the radiatively heated body of the outer ($R>10^4 R\si{G}$) disk.

The need for external illumination of the accretion disk is an important constraint on models aiming to explain the existence of double-peaked Balmer lines in AGNs. We can use the ratio of the H$\alpha$ luminosity to the viscous power available locally in the line-emitting region of the disk, $L\si{H$\alpha$}/W\si{d}$, in comparison with the \citet{Williams80} computations, to estimate the need for external illumination. The viscous power available locally in the line-emitting region of the disk, $W\si{d}$, can be estimated using an estimate of the bolometric luminosity and the extent of the emission-line region as described in, e.g., E03 and S06. In this work we use the bolometric luminosity estimates based on the observed spectral energy distributions (SEDs) presented in \S~\ref{bollum}. 
\begin{figure}
\plotone{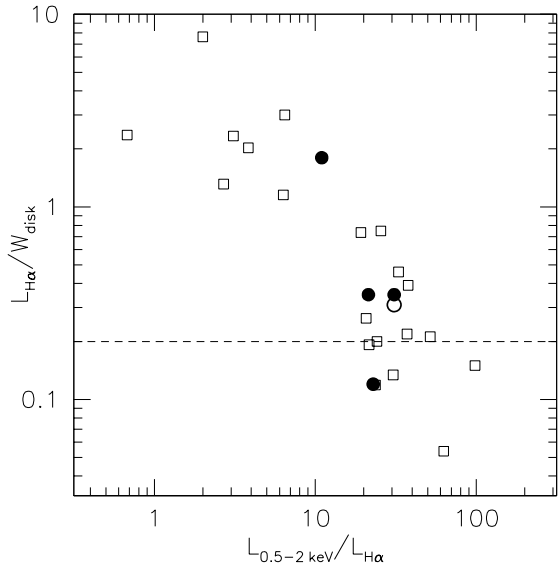}
\caption{The ratio of the X-ray flux to the H$\alpha$ line luminosity, $L\si{H$\alpha$}$,  vs. the ratio of  $L\si{H$\alpha$}$ to the viscous power released locally by the accretion disk, $W\si{d}$. The open squares denote the RL sample of E03, and the solid (open) circles indicate the RQ (RL) \emph{Chandra} double-peaked emitters. The dashed line, at $L\si{H$\alpha$}/W\si{d}=0.2$, corresponds to the \citet{Williams80} prediction for H$\alpha$-line cooling in a homogeneous, optically-thin LTE disk (see \S~\ref{diskP} for details); for objects lying above this line external illumination is necessary to produce H$\alpha$ emission of the observed strength. The apparent anti-correlation seen in this figure is just a consequence of the fact that  $W\si{d}$ is derived from the bolometric luminosity which correlates with the X-ray flux.
\label{Power}}
\end{figure}

E03 and S06 have found that for most double-peaked emitters $L\si{H$\alpha$}/W\si{d}$ ranges between 0.1 and 3. The $L\si{H$\alpha$}/W\si{d}$ ratios for the five double-peaked emitters studied here are shown in Figure~\ref{Power} in comparison to the 20 double-peaked emitters from Table~7 of E03.\footnote{We do not show the corresponding ratios for the S06 objects as the estimates were based on limits of the inner and our radii of the emitting regions rather than exact profile fits.}  Note that the apparent anti-correlation seen in this figure is just a consequence of the fact that  $W\si{d}$ is derived from the bolometric luminosity which correlates with the X-ray flux. For $L\si{H$\alpha$}/W\si{d}\gtrsim1$, the energy released locally by viscous dissipation cannot power the observed line even in principle. This is the case for one of our \emph{Chandra} targets, SDSS\,J1004$+$4801, and 7 out of the 20 (35\%) double-peaked emitters presented by E03.  Three of the remaining four \emph{Chandra} double-peaked emitters (as well as 9 out of the 20 E03 sources) have $0.2 \lesssim L\si{H$\alpha$}/W\si{d} \lesssim 1$. Thus, the observed H$\alpha$ luminosities  are uncomfortably close to the total energy available locally, and larger than the 20\% upper limit of \citet{Williams80}, suggesting that external illumination plays a role. 
The only exception is the most luminous SDSS double-peaked emitter, SDSS\,J2125$-$0813; for a standard H$\alpha$ to H$\beta$ luminosity ratio of 3.2, the measured  $L\si{H$\beta$}/W\si{d}=0.04$ corresponds to  $L\si{H$\alpha$}/W\si{d}\approx0.12$, and no external illumination is necessary.

The total soft-band X-ray luminosity is a factor of 10--30 higher than the H$\alpha$ luminosities for all five double-peaked emitters (and ranges between ~0.7 and 100 for the E03 sources), showing that enough energy is released within a few $R\si{G}$ of the black hole to illuminate the Balmer-line-emitting portion of the disk and produce line-emission with the observed strength. 

Assuming that the disk is photoionized and a thin skin emits the lines, the H$\alpha$ power depends on the illuminating flux and the reprocessing efficiency (i.e., what fraction of the cooling comes out as H$\alpha$). \citet{CD89} show that, in the case of a photoionized disk, about 20\% of the ``effective continuum power", defined as $L\si{E}=\nu L_{\nu} (3646\textrm{\,\AA})+\nu L_{\nu} (10\textrm{\,keV})$, could be emitted as H$\alpha$. In all five cases studied here, the H$\alpha$-line luminosities are equal (in the case of SDSS\,J1238$+$5325) or much smaller (by factors of 7, 6, 16, and 66, respectively, for the remaining four cases in order of increasing RA) than 20\% of the effective continuum power. 

We conclude that external illumination is necessary in at least four of the five cases studied here, since the locally available viscous energy is insufficient and the Balmer-line decrements too steep, compared to the \citet{Williams80} predictions. A photoionized disk as described in e.g., \citet{CD89}, can produce H$\alpha$-line emission of the observed strength with the proper Balmer-line decrements.

\subsection{Intrinsic Absorption}

Based on the X-ray spectral fits presented in \S~\ref{analysis}, large column density neutral absorbers are ruled out in all cases. As a simple test for the presence of ionized absorbers, we added one or more edges at $\sim0.7-0.8$\,keV to all model fits; this failed to improve the fits at significance levels $>90$\% in all cases, with the exception of SDSS\,J0132$-$0952. SDSS\,J0132$-$0952 (see \S~\ref{spec6770}), requires ionized intrinsic absorption with $N\si{H,i} \approx 4\times10^{22}$\,cm$^{-2}$ and $\xi \approx160$\,erg\,s$^{-1}$\,cm, but both quantities are poorly constrained. Silimarly, although not required by the data, the limited signal-to-noise ratio in the case of SDSS\,J2125$-$0813 allows for the addition of a $N\si{H,i}\sim7\times10^{22}$\,cm$^{-2}$, $\xi \sim 200$\,erg\,s$^{-1}$\,cm ionized intrinsic absorber.

Recent work by \citet{jet} casts doubt on the suitability of a RIAF to illuminate efficiently the outer portions of the accretion disk, suggesting instead that the required illumination can be provided by low Lorentz factor ($<2$) jets in RL objects or slow outflows in the case of RQ double-peaked emitters. If slow-moving outflows in the immediate vicinity of the black holes are responsible for the disk illumination, then X-ray observations are expected to show large X-ray absorbing columns: $N\si{H,i}\gtrsim10^{23}$\,cm$^{-2}$, providing a suitable scattering depth of $\tau\sim0.2$, according to \citet{jet}. As an example of the prevalence of suitable ionized absorbers in AGNs, \citet{jet} cite the work of \citet{Piconcelli05}, who found that 20 out of the 45 AGNs in their sample show signs of ionized absorption (parameterized by $0.7-0.8$\,keV absorption edges in 16 of the 20 cases with $\tau>0.1$, and XSPEC \emph{absori} fits in the remaining 4).

The \hbox{0.3--10\,keV} X-ray spectra of the four RQ double-peaked emitters studied here do not require large column density ionized absorbers, with the exception of SDSS\,J0132$-$0952, where a lower column-density absorber is present, and SDSS\,J2125$-$0813, where a high column-density and ionization absorber could not be ruled out. The \emph{Chandra} snapshot survey presented here has yielded spectra with a limited numbers of counts and was not designed to provide stringent constraints on the column densities and ionization states of possible ionized absorbers, but the presence of such absorbers appears unlikely in at least two of the four RQ objects from the current sample.

If high column density absorbers were common in double-peaked emitters as a class, we might also expect that the UV-to-X-ray indices measured based on X-ray photometry (e.g., S06) should reflect this by showing steeper $\alpha\si{ox}$ distributions (since the measured X-ray fluxes would be depressed by absorption). In fact, as argued above,  the double-peaked emitters tend to have flatter $\alpha\si{ox}$ values than similar optical/UV luminosity AGNs, suggesting excess X-ray emission relative to the UV emission.

We conclude that the slow-outflow model of \citet{jet} could provide the illumination necessary to power the strong double-peaked lines observed in some double-peaked emitters, but is unlikely to be responsible for the necessary extra power in all cases. 

\section{Summary}
The five double-peaked emitters observed with \emph{Chandra} show diverse properties in the X-ray band. Their  \hbox{0.3--10}\,keV emission is well fit by simple power laws; the power-law emission is modified by intrinsic absorption in the case of SDSS\,J0132$-$0952. The measured power-law photon indices, $1.6<\Gamma<1.8$, are typical for similar luminosity AGNs, with the exception of the most optically luminous object, SDSS\,J2125$-$0813. The  \hbox{0.3--10}\,keV spectrum of SDSS\,J2125$-$0813 is either absorbed, or has an unusually flat slope of $\Gamma\sim1$, and possible \ion{Fe}{25} K$\alpha$ emission. 

The soft X-ray emission is strongly  variable on timescales of years, but no short term (minutes) variability is detected, as expected for snapshot observations of similar X-ray luminosity AGNs. The optical fluxes and H$\alpha$-line fluxes also vary strongly on long timescales, with stronger variability amplitudes than typical AGN observed in the optical. 

The four RQ double-peaked emitters have strong X-ray emission (with about 30\% of the total flux emitted in the \hbox{0.5--10}\,keV band) relative to their estimated monochromatic luminosities at 2500\,\AA, and statistically flatter UV-to-X-ray spectral indices. They all show a classical BBB in the extreme UV, suggesting the absence of RIAFs. External illumination of the accretion disk is necessary to produce Balmer-line emission of the observed strength in at least four of the five cases (with the exception of the most luminous SDSS double-peaked emitter SDSS\,J2125$-$0813), and the excess X-ray emission could supply the extra energy although it is still unclear how the accretion disk will intercept a sizable fraction of the soft X-ray emission. 


\acknowledgements

IVS and WNB  acknowledge the support of NASA LTSA grant NAG5-13035.

IVS thanks Stefanie Komossa, Kazushi Iwasawa, and Andrea Merloni for helpful discussions and Jose Ramirez for his guidance using the XSTAR models.

The Hobby-Eberly Telescope (HET) is a joint project of the University of Texas at Austin, the Pennsylvania State University, Stanford University, Ludwig-Maximillians-Universit\"at M\"unchen, and Georg-August-Universit\"at G\"ottingen. The HET is named in honor of its principal benefactors, William P. Hobby and Robert E. Eberly.

The Marcario Low-Resolution Spectrograph is named for Mike Marcario of High Lonesome Optics, who fabricated several optics for the instrument but died before its completion; it is a joint project of the Hobby-Eberly Telescope partnership and the Instituto de Astronom\'{\i}a de la Universidad Nacional Aut\'onoma de M\'exico.



\begin{thebibliography}{DUM} 
\bibitem[Abramowicz et al.(2002)]{cdaf} Abramowicz, M.~A., Igumenshchev, I.~V., Quataert, E., \& Narayan, R.\ 2002, \apj, 565, 1101 
\bibitem[Almaini et al.(2000)]{Almani} Almaini, O., et al.\ 2000, \mnras, 315, 325 
\bibitem[Bianchi et al.(2007)]{Bianchi} Bianchi, S., Guainazzi, M., Matt, G., \& Fonseca Bonilla, N.\ 2007, \aap, 467, L19 
\bibitem[Brandt et al.(1997)]{Brandt97} Brandt, W.~N., Mathur, S., \& Elvis, M.\ 1997, \mnras, 285, L25 
\bibitem[Cao \& Wang(2006)]{jet} Cao, X., \& Wang, T.-G.\ 2006, \apj, 652, 112
\bibitem[Cardelli et al.(1989)]{Cardelli} Cardelli, J.~A., Clayton, G.~C., \& Mathis, J.~S.\ 1989, \apj, 345, 245 
\bibitem[Chary(2007)]{ArpSpitzer} Chary, R.-R.\ 2007, The Central Engine of Active Galactic Nuclei, 373, 443 
\bibitem[Chen et al.(1989)]{CH89a} Chen, K., Halpern, J.~P., \& Filippenko, A.~V.\ 1989, \apj, 339, 742
\bibitem[Chen \& Halpern(1989)]{CH89b} Chen, K., \& Halpern, J.~P.\ 1989, \apj, 344, 115 
\bibitem[Collin-Souffrin(1987)]{CS87} Collin-Souffrin, S.\ 1987, \aap, 179, 60 
\bibitem[Collin-Souffrin \& Dumont(1989)]{CD89} Collin-Souffrin, S., \& Dumont, A.~M.\ 1989, \aap, 213, 29 
\bibitem[Done et al.(1992)]{Done92} Done, C., Mulchaey, J.~S., Mushotzky, R.~F., \& Arnaud, K.~A.\ 1992, \apj, 395, 275
\bibitem[Dumont \& Collin-Souffrin(1990)]{dumont90} Dumont, A.~M., \& Collin-Souffrin, S.\ 1990, \aap, 229, 313 
\bibitem[Elvis et al.(1994)]{Elvis} Elvis, M., et al.\ 1994, \apjs, 95, 1 
\bibitem[Eracleous \& Halpern(1994)]{E94} Eracleous, M., \& Halpern, J.~P.\ 1994, \apjs, 90, 1
\bibitem[Eracleous et al.(1995)]{E95} Eracleous, M., Livio, M., Halpern, J.~P., \& Storchi-Bergmann, T.\ 1995, \apj, 438, 610 
\bibitem[Eracleous et al.(1996)]{2Pxray1} Eracleous, M., Halpern, J.~P., \& Livio, M.\ 1996, \apj, 459, 89 
\bibitem[Eracleous \& Halpern(1998)]{2Pxray2} Eracleous, M., \& Halpern, J.~P.\ 1998, \apj, 505, 577 
\bibitem[Eracleous et al.(2000)]{2Pxray3} Eracleous, M., Sambruna, R., \& Mushotzky, R.~F.\ 2000, \apj, 537, 654 
\bibitem[Eracleous et al.(2002)]{2Pxray4} Eracleous, M., Shields, J.~C., Chartas, G., \& Moran, E.~C.\ 2002, \apj, 565, 108 
\bibitem[Eracleous et al.(2003a)]{ArpASCA} Eracleous, M., Halpern, J.~P., \& Charlton, J.~C.\ 2003, \apj, 582, 633 
\bibitem[Eracleous \& Halpern(2003b)]{E03} Eracleous, M., \& Halpern, J.~P.\ 2003, \apj, 599, 886 
\bibitem[Eracleous et al.(2006)]{var1} Eracleous, M., Lewis, K.~T., Bogdanovi{\'c}, T., Gezari, S., \& Halpern, J.~P.\ 2006, Astronomical Society of the Pacific Conference Series, 360, 227 
\bibitem[Fioc \& Rocca-Volmerange(1997)]{FRV97} Fioc, M., \& Rocca-Volmerange, B.\ 1997, \aap, 326, 950 
\bibitem[Gezari et al.(2007)]{var2} Gezari, S., Halpern, J.~P., \& Eracleous, M.\ 2007, \apjs, 169, 167 
\bibitem[Gil de Paz et al.(2007)]{dePaz} Gil de Paz, A., et al.\ 2007, \apjs, 173, 185 
\bibitem[Grandi et al.(2006)]{3c382} Grandi, P., Malaguti, G., \& Fiocchi, M.\ 2006, \apj, 642, 113 
\bibitem[Halpern et al.(1996)]{Halpern96} Halpern, J.~P., Eracleous, M., Filippenko, A.~V., \& Chen, K.\ 1996, \apj, 464, 704 
\bibitem[Ivezi{\'c} et al.(2002)]{rl} Ivezi{\'c}, {\v Z}., et al.\ 2002, \aj, 124, 2364 
\bibitem[Ivezi{\'c} et al.(2004)]{sdssvar} Ivezi{\'c}, {\v Z}., et al.\ 2004, The Interplay Among Black Holes, Stars and ISM in Galactic Nuclei, 222, 525 
\bibitem[Kallman \& McCray(1982)]{K82} Kallman, T.~R., \& McCray, R.\ 1982, \apjs, 50, 263 
\bibitem[Kallman et al.(2004)]{K04} Kallman, T.~R., Palmeri, P., Bautista, M.~A., Mendoza, C., \& Krolik, J.~H.\ 2004, \apjs, 155, 675 
\bibitem[Kauffmann et al.(2003)]{K03} Kauffmann, G., et al.\ 2003, \mnras, 346, 1055 
\bibitem[Koester et al.(2007)]{cluster} Koester, B.~P., et al.\ 2007, \apj, 660, 239 
\bibitem[Laor(2007)]{Laor07} Laor, A.\ 2007, The Central Engine of Active Galactic Nuclei, 373, 384 
\bibitem[Lawrence \& Papadakis(1993)]{LP93} Lawrence, A., \& Papadakis, I.\ 1993, \apjl, 414, L85 
\bibitem[Lewis \& Eracleous(2006)]{Lewis} Lewis, K.~T., \& Eracleous, M.\ 2006, \apj, 642, 711
\bibitem[Magdziarz \& Zdziarski(1995)]{pexriv} Magdziarz \& Zdziarski 1995,\mnras, 273, 837
\bibitem[Mateos et al.(2007)]{Mateos} Mateos, S., Barcons, X., Carrera, F.~J., Page, M.~J., Ceballos, M.~T., Hasinger, G., \& Fabian, A.~C.\ 2007, \aap, 473, 105 
\bibitem[Narayan \& Yi(1994)]{adaf} Narayan, R., \& Yi, I.\ 1994, \apjl, 428, L13 
\bibitem[Narayan \& Yi(1995)]{NY95} Narayan, R., \& Yi, I.\ 1995, \apj, 452, 710
\bibitem[Oke(1987)]{Oke87} Oke, J.~B.\ 1987, Superluminal Radio Sources, 267 
\bibitem[Quataert \& Narayan(1999)]{adios} Quataert, E., \& Narayan, R.\ 1999, \apj, 520, 298 
\bibitem[Piconcelli et al.(2002)]{piconcelli} Piconcelli, E., Cappi, M., Bassani, L., Fiore, F., Di Cocco, G., \& Stephen, J.~B.\ 2002, \aap, 394, 835 
\bibitem[Piconcelli et al.(2005)]{Piconcelli05} Piconcelli, E., Jimenez-Bail{\'o}n, E., Guainazzi, M., Schartel, N., Rodr{\'{\i}}guez-Pascual, P.~M., \& Santos-Lle{\'o}, M.\ 2005, \aap, 432, 15 
\bibitem[Rees et al.(1982)]{Rees82} Rees, M.~J., Begelman, M.~C., Blandford, R.~D., \& Phinney, E.~S.\ 1982, \nat, 295, 17 
\bibitem[Richards et al.(2006)]{Richards06} Richards, G.~T., et al.\ 2006, \apjs, 166, 470
\bibitem[Riffel et al.(2006)]{Riffel} Riffel, R., Rodr{\'{\i}}guez-Ardila, A., \& Pastoriza, M.~G.\ 2006, \aap, 457, 61 
\bibitem[Schlegel et al.(1998)]{Schlegel} Schlegel, D.~J., Finkbeiner, D.~P., \& Davis, M.\ 1998, \apj, 500, 525 
\bibitem[Shakura \& Sunyaev(1973)]{SS73} Shakura, N.I, Sunyaev, R.A.\ 1973, Astron. \& Astrophys., 24, 337 
\bibitem[Shemmer et al.(2006)]{Shemmer06} Shemmer, O., Brandt, W.~N., Netzer, H., Maiolino, R., \& Kaspi, S.\ 2006, \apjl, 646, L29 
\bibitem[Shemmer et al.(2008)]{Shemmer08} Shemmer, O., Brandt, W.~N., Netzer, H., Maiolino, R., \& Kaspi, S.\ 2008, ArXiv e-prints, 804, arXiv:0804.0803 
\bibitem[Shinozaki et al.(2006)]{Shinozaki} Shinozaki, K., Miyaji, T., Ishisaki, Y., Ueda, Y., \& Ogasaka, Y.\ 2006, \aj, 131, 2843 
\bibitem[Spergel et al.(2007)]{Spergel07} Spergel, D.~N., et al.\ 2007, \apjs, 170, 377 
\bibitem[Steffen et al.(2006)]{St06} Steffen, A.~T., Strateva, I., Brandt, W.~N., Alexander, D.~M., Koekemoer, A.~M., Lehmer, B.~D., Schneider, D.~P., \& Vignali, C.\ 2006, \aj, 131, 2826 
\bibitem[Strateva et al.(2006)]{S06} Strateva, I.~V., Brandt, W.~N., Eracleous, M., Schneider, D.~P., \& Chartas, G.\ 2006, \apj, 651, 749 
\bibitem[Strateva et al.(2007)]{S07} Strateva, I., Brandt, W.~N., Eracleous, M., Garmire, G., \& Komossa, S.\ 2007, The Central Engine of Active Galactic Nuclei, 373, 399 
\bibitem[Strateva et al.(2003)]{S03} Strateva, I.~V., et al.\ 2003, \aj, 126, 1720
\bibitem[Nandra et al.(1997)]{Nandra97} Nandra, K., George, I.~M., Mushotzky, R.~F., Turner, T.~J., \& Yaqoob, T.\ 1997, \apj, 476, 70  
\bibitem[Tarter et al.(1969)]{T69} Tarter, C.~B., Tucker, W.~H., \& Salpeter, E.~E.\ 1969, \apj, 156, 943 
\bibitem[Thorne(1974)]{Thorne74} Thorne, K.~S.\ 1974, \apj, 191, 507 
\bibitem[Vanden Berk et al.(2001)]{Berk} Vanden Berk, D.~E., et al.\ 2001, \aj, 122, 549 
\bibitem[Wilkes et al.(1994)]{rlagn} Wilkes, B.~J., Tananbaum, H., Worrall, D.~M., Avni, Y., Oey, M.~S., \& Flanagan, J.\ 1994, \apjs, 92, 53 
\bibitem[Williams(1980)]{Williams80} Williams, R.~E.\ 1980, \apj, 235, 939 
\bibitem[Yip et al.(2004a)]{Yip1} Yip, C.~W., et al.\ 2004, \aj, 128, 585 
\bibitem[Yip et al.(2004b)]{Yip2} Yip, C.~W., et al.\ 2004, \aj, 128, 2603
\end{thebibliography}
\end{document}